\documentclass[fleqn,usenatbib]{mnras}
\usepackage{newtxtext,newtxmath}
\usepackage[T1]{fontenc}
\usepackage{ae,aecompl}

\usepackage{graphicx}	% Including figure files
\usepackage{amsmath}	% Advanced maths commands
\usepackage{amssymb}	% Extra maths symbols
\title[class II YSO outbursts]{Determining the recurrence timescale of long-lasting YSO outbursts}

\author[C. Contreras Pe\~{n}a et al.]{
Carlos Contreras Pe\~{n}a,$^{1}$\thanks{E-mail: c.contreras@exeter.ac.uk (CCP)}
Tim Naylor,$^{1}$
and Sam Morrell$^{1}$
\\
$^{1}$School of Physics, Astrophysics Group, University of Exeter, Stocker Road, Exeter EX4 4QL, UK
}

\date{Accepted XXX. Received YYY; in original form ZZZ}

\pubyear{2019}

\begin{document}
\label{firstpage}
\pagerange{\pageref{firstpage}--\pageref{lastpage}}
\maketitle

\begin{abstract}
We have determined the rate of large accretion events in class I and II young stellar objects (YSOs) by comparing the all-sky digitised photographic plate surveys provided by SuperCOSMOS with the latest data release from {\it Gaia} (DR2). The long mean baseline of 55 years along with a large sample of class II YSOs ($\simeq$15,000) allows us to study approximately 1 million YSO-years. We find 139 objects with $\Delta R\geq1$~mag, most of which are found at amplitudes between 1 and 3 mag. The majority of YSOs in this group show irregular variability or long-lasting fading events, which is best explained as hot spots due to accretion or by variable extinction. There is a tail of YSOs at $\Delta R\geq3$~mag and they seem to represent a different population. Surprisingly many objects in this group show high-amplitude irregular variability over timescales shorter than 10 years, in contrast with the view that high-amplitude objects always have long outbursts. However, we find 6 objects that are consistent with undergoing large, long lasting accretion events, 3 of them previously unknown. This yields an outburst recurrence timescale of 112 kyr, with a 68\% confidence interval [74 to 180] kyr. This represents the first robust determination of the outburst rate in class II YSOs and shows that YSOs in their planet-forming stage do in fact undergo large accretion events, and with timescales of $\simeq$100,000 years. In addition, we find that outbursts in the class II stage are $\simeq$10 times less frequent than during the class I stage.
\end{abstract}


%%%%%%%%%%%%%%%%%%%%%%%%%%%%%%%%%%%%%%%%%%%%%%%%%%

%%%%%%%%%%%%%%%%% BODY OF PAPER %%%%%%%%%%%%%%%%%%

\section{Introduction}\label{sec:intro}

The crucial period for planet formation is between $\simeq$1 Myr \citep[when the first planetesimals in our own solar system were formed][]{2015Pfalzner}, and $\simeq$10 Myr when the protoplanetary discs around most stars have dissipated \citep[e.g.][]{2013Bell}.
This corresponds exactly to the Class II phase of young stellar evolution \citep[see e.g.][for a definition of the stages of young stellar evolution]{2014Dunham}. Therefore properties of the disc, such as surface density or temperature, play a key role in the formation and evolution of protoplanets. These properties determine, for example, the outer boundary condition for the gas accretion rate of the protoplanets, enter into migration rates or determine the location of the snowline, the latter having an impact on the surface density of solids \citep[see e.g.][]{2015Mordasini}.  

The aforementioned disc characteristics depend on the accretion rate from the disc onto the central star. Theoretical models show that accretion onto the central star is unlikely to be a steady process \citep[see e.g.][]{2015Vorobyov}, with observational support arising from the analysis of knots along jets from young stars \citep[e.g.][]{2018Makin} as well as outbursts in YSOs \citep[e.g.][]{1996Hartmann}. However, planet formation models fail to include this effect and generally assume that the accretion rate decreases steadily with time \citep[e.g. ][]{2015Mulders}, with some models including a dependence with the mass of the central star \citep{2008Kennedy}.

The changes in the accretion rate through circumstellar discs (if they occur) could have a profound effect on the emerging planetary systems. For example \citet{2017Hubbardb} argues that large accretion events can help to solve the so-called meter barrier problem \citep[][]{2014Boley} and allow the in-situ formation of rocky planets even at distances of $\simeq$1.5 AU. It also helps to explain the small size of Mars \citep{2014Chambers}. These events can also promote the formation of Jupiter and Saturn-like gas giants \citep{2017Hubbarda}. Finally, the increase of the central luminosity due to changes of the accretion rate (as small as a factor of two) will drive the snow lines for various ices (used to predict the composition of planets) towards larger radii \citep{2007Garaud}. This effect has already been observed in the known eruptive young stellar object (YSO) V883 Ori, where the water snow line is found at a distance of 42 AU from the central star, far beyond the expected location for a 1 M$_{\odot}$  star \citep{2016Cieza}. The observations of calcium-aluminium rich inclusions in chondrites \citep{2009Wurm} and the depletion of lithophile elements in Earth \citep{2014Hubbard} could be evidence for large eruptions in our own solar system. 

Variable accretion in YSOs can be caused by a variety of different physical mechanisms and there may exist a continuum of outbursting behaviour with a range of amplitudes and timescales \citep{2017Cody}. Stochastic accretion bursts in an instability-driven accretion regime lead to changes in the accretion rate up to 700$\%$ over 1-10 days \citep[e.g.][]{2014Venuti, 2017Cody}. Large changes of accretion rate, increasing from typical rates of 10$^{-7}$~M$_{\odot}$~yr$^{-1}$ up to 10$^{-4}$~M$_{\odot}$~yr$^{-1}$, with outburst durations of weeks to 100 years, which can increase the central luminosity to 100L$_{\odot}$ \citep[e.g.][]{1996Hartmann}, are observed in eruptive YSOs \citep[the FUors, EXors and MNors, e.g.][]{2014Audard, 2017Contreras}. Here, some kind of disc instability, such as gravitational instabilities \citealp[][]{2005Vorobyov} or planet-induced thermal instabilities \citep[e.g.][]{2004Lodato}, are suggested as explanations of the large variability. 

However, many aspects of the so-called eruptive YSOs are still uncertain. The frequency and amplitude of the outbursts is not well constrained \citep[e.g.][]{2013Scholz}. In addition, there is controversy as to whether the very largest outbursts are associated with the Class II planet building phase at all, or are just limited to the pre-planet-forming (Class 0/I) phase (\citealp[c.f.][]{2001Sandell}, \citealp[with][]{2011Miller}). 

In this paper we present the results of the study we conducted using {\it Gaia} and the Schmidt photographic plate surveys for a large sample of YSOs. It is our aim to answer the questions of outburst amplitudes and frequencies during the planet-forming stage of young stellar evolution. This paper is divided as follows: in Section \ref{sec:sample} we present the data used to maximise the time baseline and the sample size as well as describing how the sample was classified as class II YSOs. In Section \ref{sec:method} we present an outline of our method to select high-amplitude variability and discuss the issues we found when applying the method to our sample. Section \ref{sec:class2amp} presents a discussion on the distribution of amplitudes for the variability of class II YSOs. In Section \ref{sec:sel_hamp} we show the steps taken to select the YSOs showing long-term outbursts where variability is most likely driven by dramatic episodes of enhanced accretion. From the latter sample, in Section \ref{sec:outrate} we determine the outburst rate during the class II stage and compare it to previous theoretical and observational estimates.  In section \ref{sec:caveats} we present a number of caveats that could affect our estimate of the outburst rate, whilst in Section \ref{sec:out_mec} we discuss the implications of our result on the likely physical mechanisms driving the outburst during the class II stage. Finally, Section \ref{sec:sum} shows a summary of our results.

%Gravitationally unstable phases in a protoplanetary disc result in large increases in accretion rate through the disc, which protoplanets can survive, but at the expense of driving their orbital evolution \citep{2013Boss}.The stability of magnetic structures at the star and inner disc can dictate the orbits of hot Jupiters \citep[e.g.][]{1996Lin}, yet we know those magnetic structures change with accretion rate.

\section{Data}\label{sec:sample}

To establish the recurrence rate of accretion related outbursts during the planet-forming stage, it is important to maximise both the time baseline and the number of YSOs surveyed \citep[see e.g.][]{2015Hillenbrand}.  In this paper we achieve both by comparing the magnitudes of a large sample of class II YSOs from the photographic atlases provided by the SuperCOSMOS Sky Survey \citep[hereafter SSS,][]{2001Hambly} with the latest {\it Gaia} data release.

We note that the current classification system of eruptive YSOs is a largely phenomenological one based on photometric and spectroscopic characteristics, with long-term outbursts falling into the FU Orionis (FUor) class and short-term, repetitive outbursts classified as EXors. However, the more recent data show characteristics that have been difficult to classify into the original sub-classes \citep[see e.g.][]{2017Contreras}. Given this, during the remainder of the paper we will not ascribe a class to the dramatic accretion events that we are searching for, but we will simply use a physical classification and refer to them as high-amplitude, long-term accretion events.

\subsection{Time baseline}

The SSS has digitised the entire sky in three colours ($B, R$ and $I$) whilst providing a second epoch in $R$, by scanning Schmidt photographic plates  \citep[][]{2001Hambly}. The observations took place during the second half of the twentieth century, reaching a depth of $R\simeq20$~mag. The photographic surveys included in SuperCOSMOS are the SERC-J/EJ \citep{1984Cannon}, SERC-ER \citep{1984Cannon}, AAO-R \citep{1992Morgan}, ESO-R \citep{1984West}, POSS-I E \citep{1963Minkowski} and POSS-II B, R and I \citep{1991Reid} surveys. For a more detailed description see table 1 of \citet{2001Hambly}.

The {\it Gaia} mission \citep{2018Gaiadr2} has provided the first all-sky photometric survey that is able to match the depth of the photographic Palomar and ESO/SERC surveys. This provides the chance of studying the long term variability of approximately 1 billion stars with a minimum and maximum time baselines of 15 and 66 years respectively. 

\subsection{The sample}

\subsubsection{Initial selection}

To obtain the largest possible sample, we assembled a very inclusive sample of YSOs, and then identified the class II YSOs using their spectral energy distribution (SED).

The first step consisted of searching the SIMBAD database \citep{2000Wenger} for Galactic sources with classifications that make them likely to be a young stellar object. We included objects with classification as pre-main sequence stars (pr*), pre-main sequence star candidate (pr?), emission lines star (em*), variable star of Orion type (or*), Herbig Ae/Be star (Ae*), Herbig Ae/Be star candidate (Ae?), Herbig-Haro object (HH), FU Orionis star (FU*), T Tauri-type star (TT*), T Taur-type star candidate (TT?), young stellar object (Y*O) and young stellar object candidate (Y*?).  The search yielded 77873 objects.

%{\bf We attempted to define a sample that was unrelated to variability. We note that objects in two SIMBAD classes, FU* and or*, have also been defined as YSOs based on characteristics other than their variability. In addition the inclusion of objects classified as FU Orionis stars was done to check whether our method was successfully retrieving long-lasting large amplitude outbursts. } 

The SIMBAD classification does not indicate the likely evolutionary stage of the objects in our sample. The observed SED of young stellar objects between 2 $< \lambda <$ 20 $\mu$m is generally used to determine the likely evolutionary stage of the system \citep[e.g.][]{1987Lada, 1994Greene, 2014Dunham}. Objects where the infalling envelope has dissipated, but the star is still accreting material from a circumstellar disc, are known as class II YSOs, and it is during this stage where planet formation is believed to occur (see Section \ref{sec:intro}). Therefore we had to determine the fraction of our sample that show colours and SEDs consistent with those of class II YSOs. 

\subsubsection{Classification}\label{sec:ysoclass}

\begin{figure*}
	\resizebox{0.95\columnwidth}{!}{\includegraphics{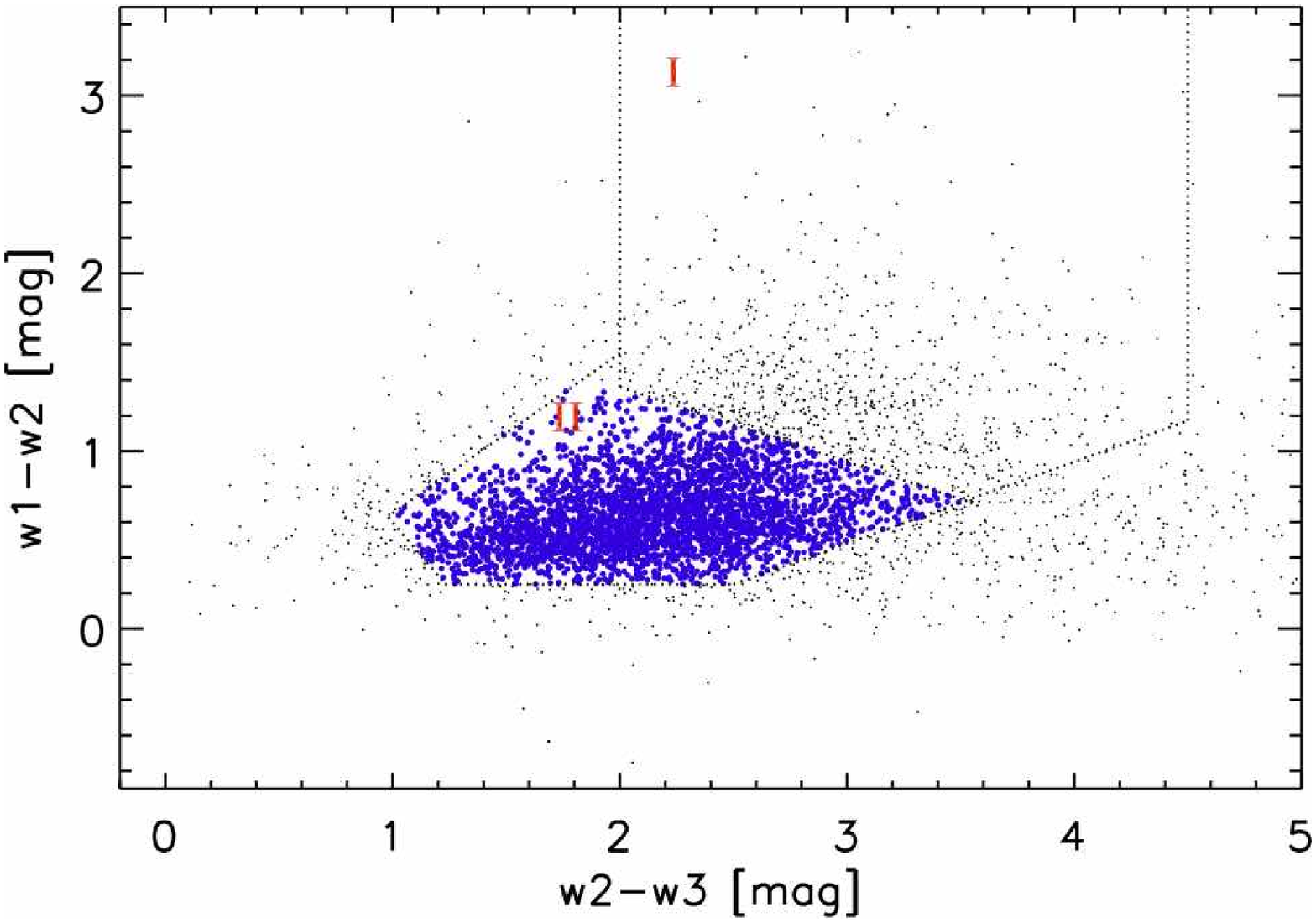}}
        \resizebox{0.95\columnwidth}{!}{\includegraphics{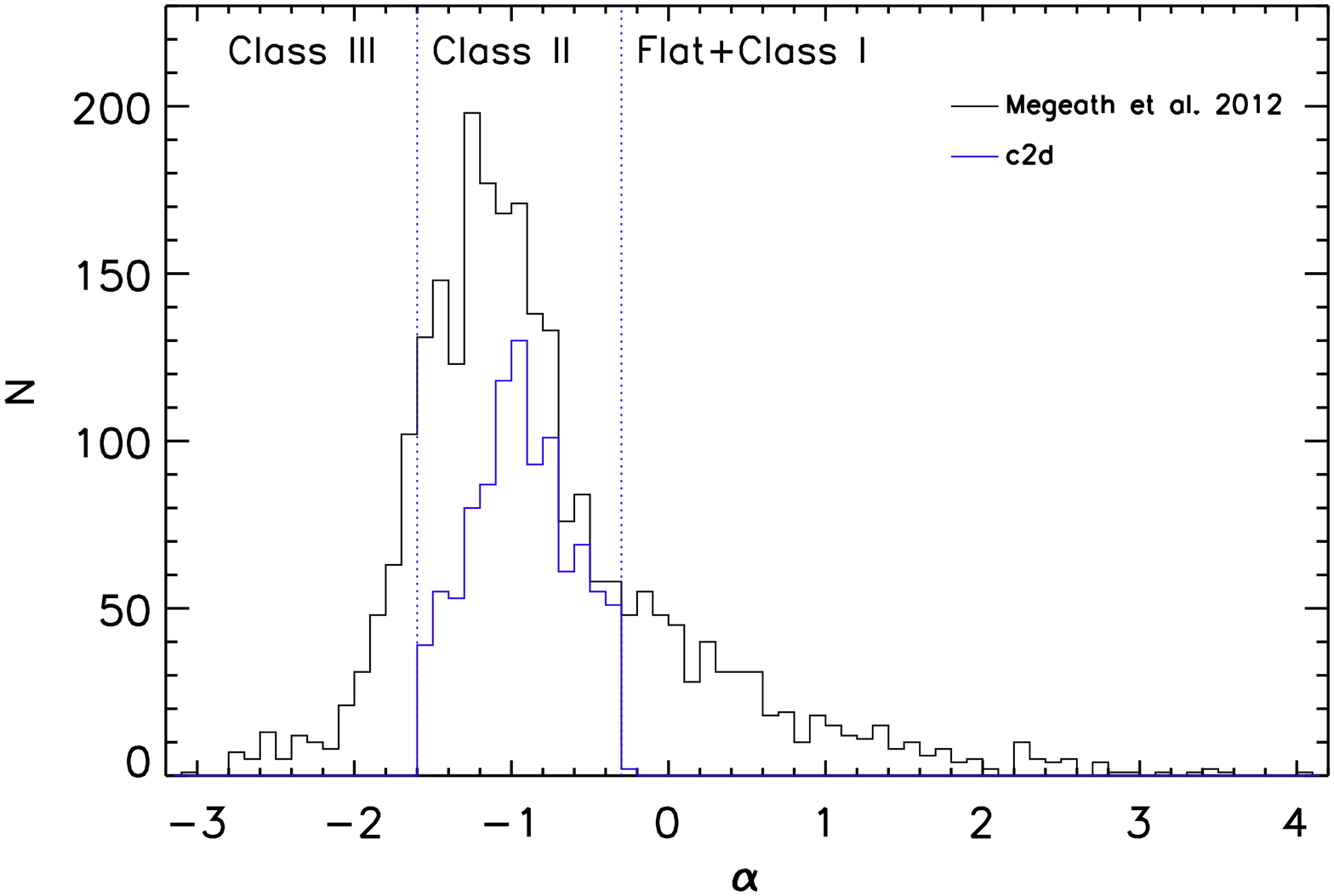}}\\
        	\resizebox{0.95\columnwidth}{!}{\includegraphics{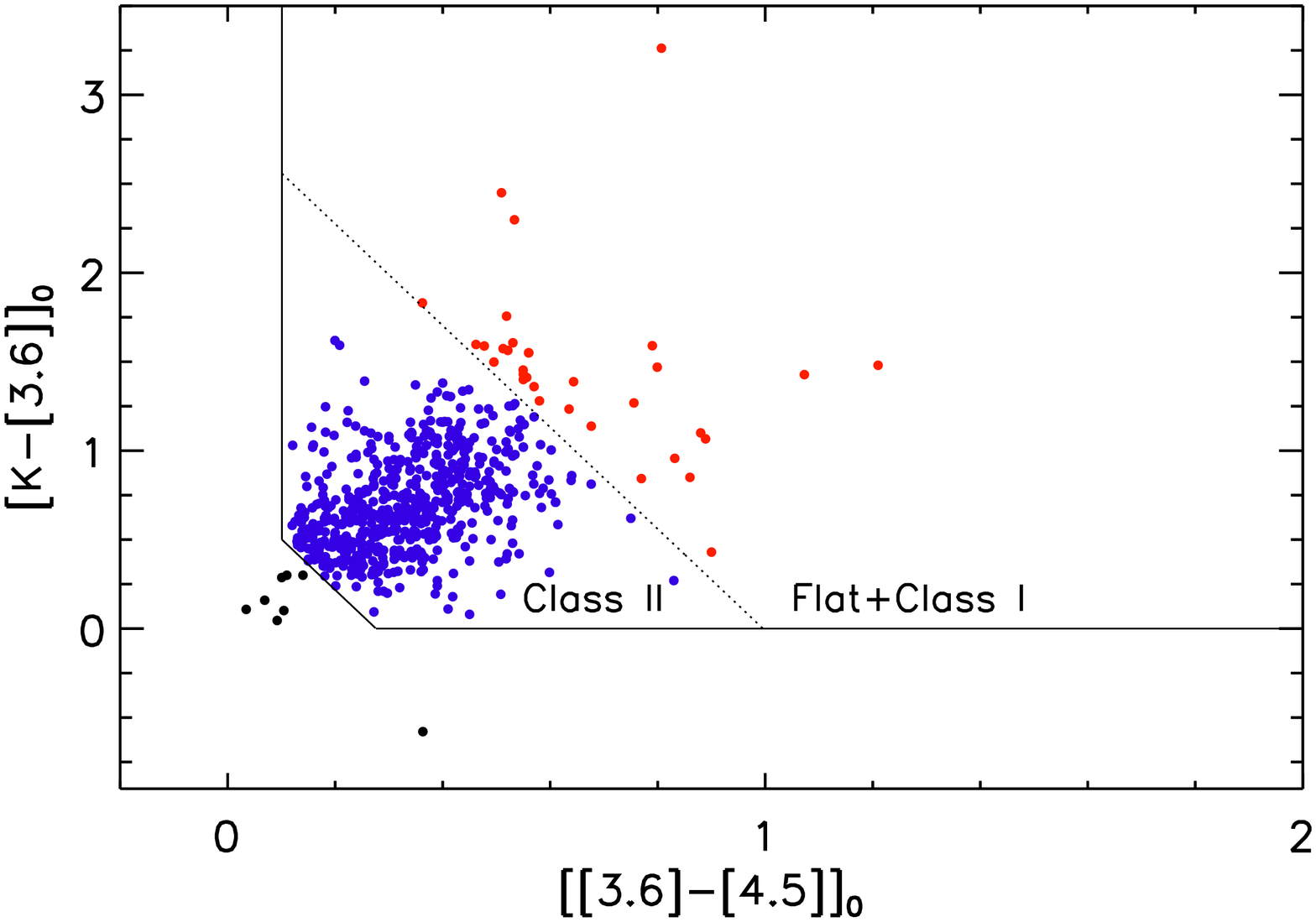}}
	\resizebox{0.95\columnwidth}{!}{\includegraphics{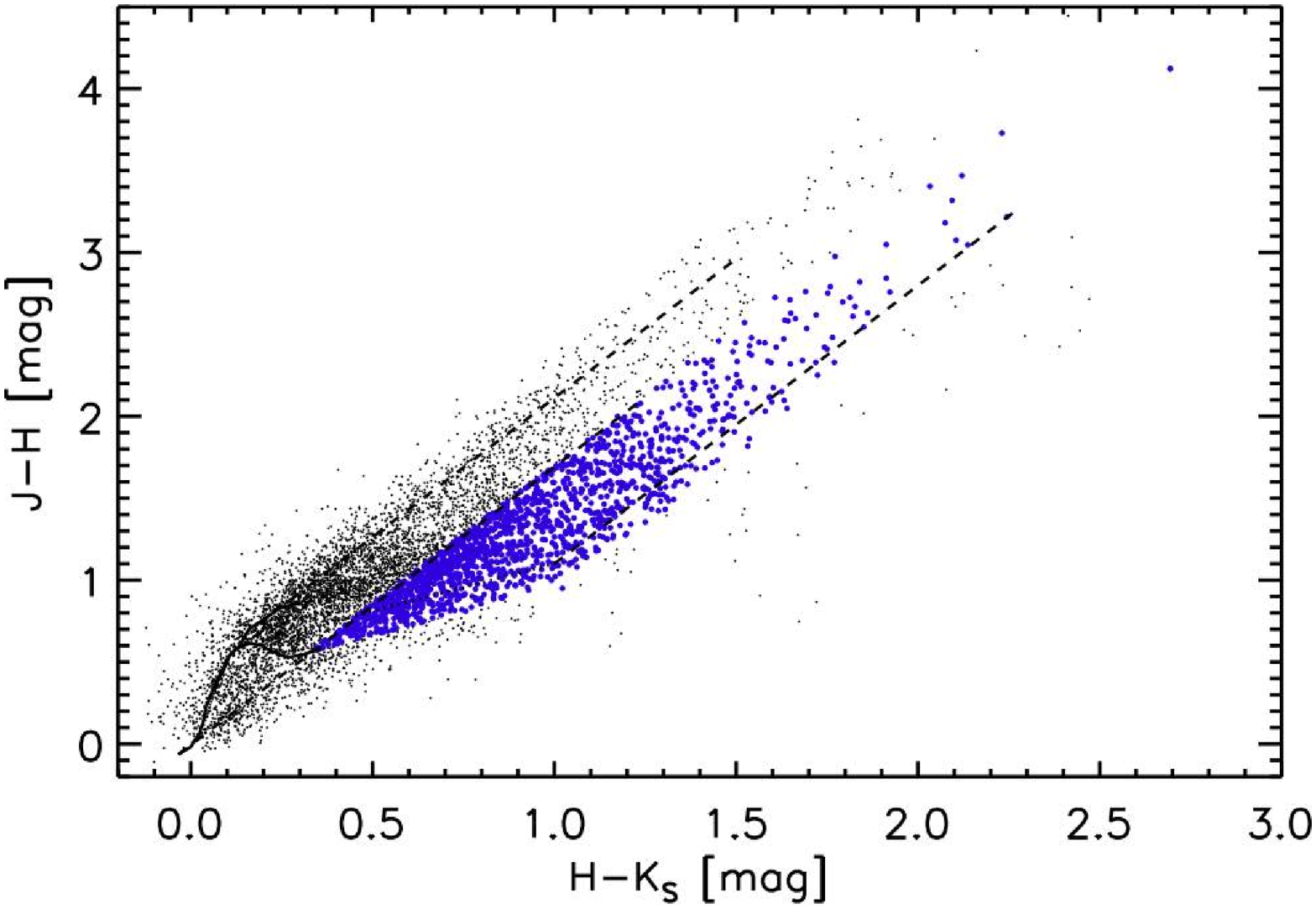}}
    \caption{Examples of the different methods used to classify young stellar objects into the different classes of young stellar evolution. (top-left) $w1-w2$ vs $w2-w3$ colour-colour plot for objects found in the \citet{2016Marton} catalogue (small black dots). The dotted lines mark the regions for class I and class II YSOs according to the \citet{2014Koenig} criteria. In the figure we mark objects that fall in the class II regions with blue circles. (top-right) YSOs found in \citet{2012Megeath} (solid black line) which are classified based on the spectral index determined from {\it Spitzer} photometry. In addition the distribution of c2d class II YSOs determined from $\alpha$ is shown in the solid blue line. The values that limit the class definitions are shown in blue dotted lines. (bottom-left) YSOs found in \citet{2012Megeath} for which no value of $\alpha$ is published but possess reliable K$_{\rm s}$, I1 and I2 photometry (see main text). In the plot class I YSOs are shown with red circles, class II YSOs in blue circles and unclassified objects are marked with black circles. (bottom-right) $J-H$ vs $H-K_{\rm s}$ colour-colour diagram for sources that lack other forms of classification and that have reliable 2MASS photometry. Class II YSOs are marked in blue circles.}
    \label{fig:class}
\end{figure*}

We crossmatched our sample (r$<$1\arcsec) with several studies that provide a class for YSOs in known star formation regions\footnote{We note that we did not attempt to search for every existing catalogue that provides a YSO classification. However, using these catalogues we were still able to create a large sample of class II YSOs.} \citep{2003Evans,2008Chavarria, 2008Connelley, 2008Gutermuth, 2008Koenig, 2009Evans, 2009Gutermuth, 2009Kirk, 2010Billot, 2010Rebull, 2011Rebull, 2011Rivera, 2012Allen, 2012Megeath, 2016Marton}. All of these provide a classification based on the near- to mid-IR SEDs of the YSOs. We found that most of the objects with a crossmatch were contained within the studies of \citet{2016Marton}, \citet{2012Megeath}, \citet{2003Evans} and \citet{2009Gutermuth}. The \citeauthor{2016Marton} study is based on the all-sky survey {\it WISE} and as such contains the largest  number of sources. Given this, for better consistency when trying to classify our sample we gave the \citet{2016Marton} study preference over \citeauthor{2012Megeath},  \citeauthor{2003Evans} and \citeauthor{2009Gutermuth} as these are based on specific areas of star formation.  The classification of YSOs in our sample is explained below.

\begin{description}

\item {\bf \citet{2016Marton}}. Through the use of a supervised learning algorithm \citet{2016Marton} classify objects as YSOs based on 2MASS $JHK_{\rm s}$ and {\it WISE} \citep{2010Wright} photometry. However, the study 
does not provide a direct classification as class II YSOs, and instead yields a classification either as class I/II or class III YSOs. To identify the class II YSOs, we made use of the classification criteria of \citet{2014Koenig} which are based on {\it WISE} bands $W1$ (3.4 $\mu$m), $W2$ (4.6 $\mu$m) and $W3$ (12 $\mu$m) bands (see Fig. \ref{fig:class}). The latter yielded 3253 class II YSOs. We note that \citet{2014Koenig} define different criteria to discard possible contaminant sources (AGNs, resolved PAH emission, among others) before attempting to classify sources in their sample as YSOs, however pre-selection using the \citeauthor{2016Marton} catalogue has ensured our sample comprises true YSOs and as such we do not believe the criteria select contaminant objects.  

\item {\bf \citet{2012Megeath}}. This study is based on 2MASS $JHK_{\rm s}$ \citep{2006Skrutskie}, IRAC \citep{2004Fazio} and MIPS \citep{2004Rieke} {\it Spitzer} photometry of the Orion A and B molecular clouds. The authors provide a value for the slope of the SED at the IRAC wavelengths (spectral index, $\alpha_{\mathrm{IRAC}}$). From the latter, 1671 YSOs show values -1.6~$< \alpha<$~-0.3 (see Fig. \ref{fig:class}); following \citet{1994Greene} these were classified as class II YSOs. From those objects with no value of $\alpha$, 605 have reliable $K_{s}$, I1 (3.6 $\mu$m) and I2 (4.5 $\mu$m) photometry as well as values for the $K$-band extinction, $A_{Ks}$.  We were thus able to classify them as class II YSOs using the criteria established in Appendix A of \citet{2009Gutermuth} (see Fig. \ref{fig:class}). For the remaining objects which cannot be classified through these two methods, we found 86 objects that are classified as disc objects in \citet{2012Megeath}, thus we assumed that these were very likely class II YSOs.

\item {\bf \citet{2009Gutermuth}}. The authors provide 2MASS $JHK_{\rm s}$, IRAC and MIPS {\it Spitzer} photometry for 36 young, nearby star-forming clusters, where they also provide an evolutionary class for YSOs in these areas (see their Appendix A). Based on this classification, we found 1060 class II YSOs that were added to our list. 

\item {\bf \citet{2003Evans}}. The ``Cores to discs'' (c2d) {\it Spitzer} legacy project provides 2MASS $JHK_{\rm s}$, IRAC and MIPS {\it Spitzer} photometry for five large and nearby molecular clouds \citep[see e.g][]{2009Evans}. The project also provides the slope of the SED, $\alpha$, using the region between 2 and 24 $\mu$m. Using the latter information we added 994 class II YSOs to our list.

\end{description}

For the objects in the SIMBAD list that were not found in any of the catalogues mentioned above, we established a likely class through the following steps.

\begin{description}

\item  {\it {\bf WISE}}.  We used the ALLWISE data release \citep{2013Cutri}. We found 4883 objects that were detected in the $W1, W2$ and $W3$ bands with uncertainties below 0.4\footnote{We note that \citet{2014Koenig} require $\sigma<0.2$ in al three bands, however we used a less strict criterion in order to include more objects in our classification.}~mag in all three bands and that can be flagged as class II YSOs (see Fig. \ref{fig:class}) using the criteria of \citet{2014Koenig}.

\item {\bf 2MASS}. 1536 objects have reliable 2MASS photometry (quality flag ``AAA'') and can be flagged as class II YSOs based on their location in the $J-H$ vs $H-K_{\rm s}$ colour-colour diagram (see Figure \ref{fig:class}), i.e. they show larger $H-K_{\rm s}$ colours than main-sequence stars and have colours that are consistent with those of classical T Tauri stars (CTTS) as defined by the CTTS locus of \citet{1997Meyer}. 

\item{\bf MYStIX}. Many YSOs found in the Massive Young Star-Forming Complex Study in Infrared and X-ray \citep[MYStIX][]{2013Feigelson} project are not part of the SIMBAD database. We found that 1400 of these were classified as being in the planet-forming stage, thus they were also added to our sample.

\end{description}

In summary we were able to compile a list of 15404 class II YSOs.

%If we consider a mean epoch difference between Gaia and SuperCOSMOS of 35 years, then we have effectively covered $\simeq$535000 YSO years of the planet-forming stage.

\section{{\it Gaia} vs SuperCOSMOS search}\label{sec:method}

\subsection{Outline method}

To search for high-amplitude variables in our sample we compared their {\it Gaia} magnitudes with those found in each individual SuperCOSMOS band. The SSS data were obtained from the merged source catalogue of SSS via a structured data language (SQL) query of the SuperCOSMOS Science Archive (SSA\footnote{http://ssa.roe.ac.uk/}), whilst {\it Gaia} magnitudes were obtained from the second data release (DR2) catalogue downloaded to a local machine. The method was as follows.

\begin{enumerate}

\item  For each YSO we searched for a {\it Gaia} and SSS counterpart, using a separation of 1.6\arcsec and 2\arcsec respectively. If the YSO was not found in {\it Gaia}, then we set the magnitude of the object to be $G=21$ to account for the possibility that the star was fainter than the {\it Gaia} limit at the time of observations. Whenever the YSO was not detected in an SSS band, we set the magnitude of the YSO to be that of the faintest SSS source with a stellar profile (class 2) in a $2^{\circ}\times2^{\circ}$ box around the YSO in the corresponding band.   

\item We crossmatched {\it Gaia} and SSS sources found in a $2^{\circ}\times2^{\circ}$ box around each YSO, where SSS detections with stellar classification were selected. This allowed us to determine the relation between  the {\it Gaia} $G$ broadband magnitude and the $B$, $R1$, $R2$ and $I$ magnitudes obtained from SSS. 

\item For each SSS band (whose corresponding magnitude  we will call $M_{\rm sss}$), we determined $G-M_{\rm sss}$, for all objects with {\it Gaia} magnitudes that were found within $|G-G_{\rm YSO}|\leq0.2$, where $G_{\rm YSO}$ is the {\it Gaia} magnitude of the YSO being analysed. The mean value of the $G-M_{\rm sss}$ distribution gave us the expected difference between {\it Gaia} and SSS for (what we assume are largely) non-variable stellar sources in the region around the YSO. A large difference between the observed $G-M_{\rm sss}$ colour of the YSO vs the expected $G-M_{\rm sss}$ colour estimated from non-variable sources, gave us an indication as to whether the YSO had shown any variability between the photographic surveys and {\it Gaia} DR2.

We quantified the variability of the YSO by treating the $G-M_{\rm sss}$ distribution as a 4-dimensional multivariate normal distribution. We define an statistic for the variability of the source, $\Delta_{\mathrm{YSO}}$, as

\begin{equation}
    \Delta_{\mathrm{YSO}}=(\boldsymbol{x}-\boldsymbol{u})^{T}\boldsymbol{\Sigma}^{-1}(\boldsymbol{x}-\boldsymbol{u})-\chi^{2}(5\sigma)_{k=4},
\end{equation}

\noindent with $\boldsymbol{x}$ the vector of the observed $G-M_{\rm sss}$ colours for the YSO, $\boldsymbol{u}$ the vector of expected $G-M_{\rm sss}$ colours estimated for non-variable stars, $\Sigma$ is the covariance matrix determined from the expected $G-M_{\rm sss}$ colours, and $\chi^{2}(5\sigma)_{k=4}$ is the value of the $\chi^{2}$ distribution with 4 degrees of freedom, and for a confidence interval of 99.99996$\%$ or 5$\sigma$.

\begin{figure*}
	\resizebox{\columnwidth}{!}{\includegraphics{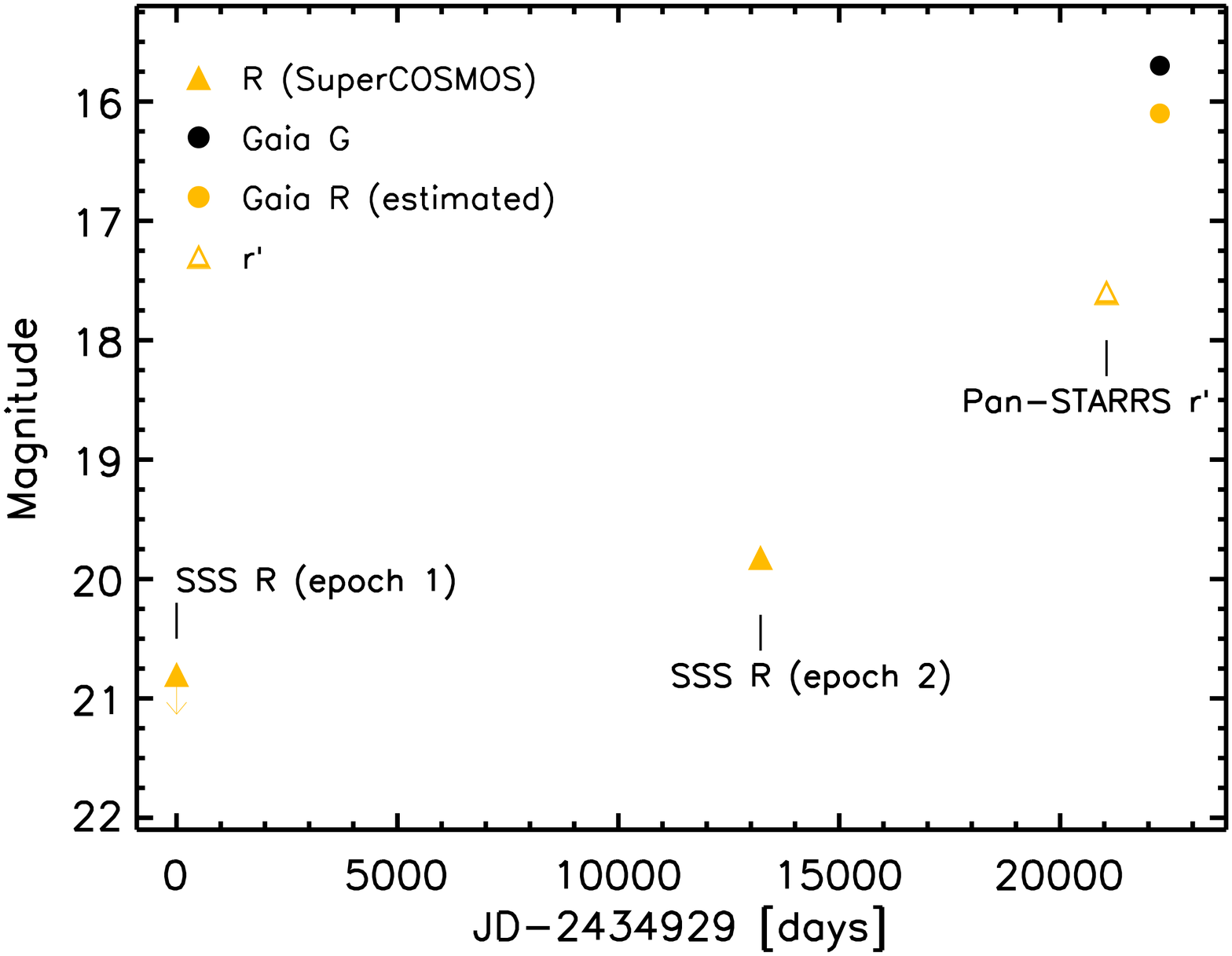}}
	\resizebox{\columnwidth}{!}{\includegraphics{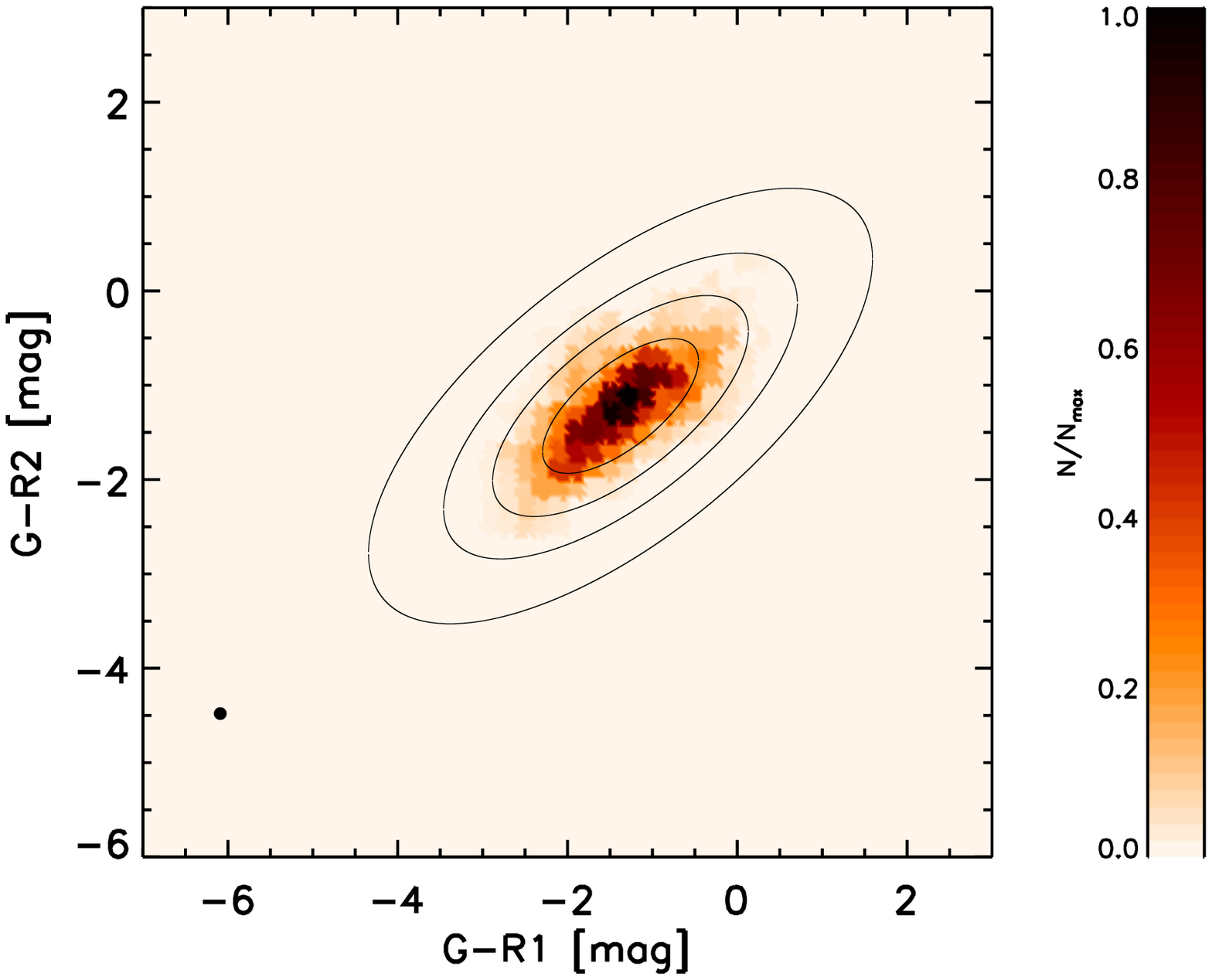}}
    \caption{(left) Light curve of V2492 Cyg including information from SuperCOSMOS, Pan-STARRS and the latest magnitude provided by {\it Gaia}. In the plot we mark the approximate epoch of each survey for which we present images in Figure \ref{fig:v2492Cyg_a}. (right) $G-R$ (2nd epoch) vs $G-R$ (1st epoch) distribution for objects found in a 2$\times2^{\circ}$ box around V2492 Cyg. 1, 2, 4 and 5$\sigma$ confidence intervals are shown as thick black lines. The location of V2492 Cyg is shown as a black circle. }
    \label{fig:v2492Cyg_b}
\end{figure*}

To illustrate the type of objects we were looking for, we analysed the known eruptive YSOs that were part of the {\it Gaia} alerts sample, V2492 Cyg \citep[e.g.][]{2013Hillenbrand}, V350 Cep \citep{1999Magakian} and ASASSN13-db \citep{2017Sicilia}. In Figure \ref{fig:v2492Cyg_a} we show the example of eruptive class I YSO V2492 Cyg. The YSO is not detected in the first epoch of $R$ (observations on September 1952), is faint at the second epoch (observations on July 1989) with $R=19.8$~mag and is bright during {\it Gaia} observations ($G\simeq15$~mag). Figure \ref{fig:v2492Cyg_b} shows a 2-dimensional representation of our method applied to V2492 Cyg, where it can be seen that the observed $G-R1$ vs $G-R2$ colours of the object are located well beyond the 5$\sigma$ confidence intervals. We find similar results for V350 Cep and ASASSN13-db.

\subsection{Finding a suitable $\Delta_{\mathrm{YSO}}$}\label{sec:mt}

We began by classifying the YSO as a variable star candidate if it satisfied the following condition

\begin{equation}
    \Delta_{\mathrm{YSO}} > 0,
	\label{eq:varcon}
\end{equation}

\noindent which ensures the selection of high-amplitude variable stars.

\end{enumerate}

\begin{figure*}
	\resizebox{0.95\textwidth}{!}{\includegraphics{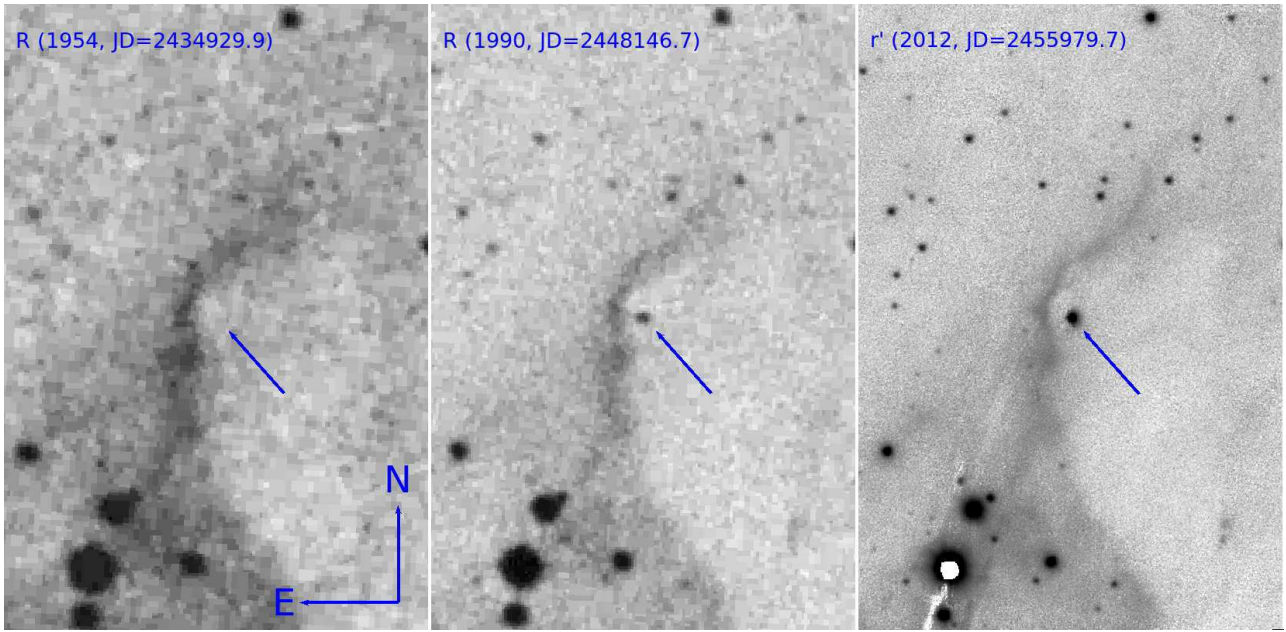}}
    \caption{SuperCOSMOS POSS-I E (1954) and POSS-II R (1990) $R$ plate images, and Pan-STARRS $r^{'}$ (2012) image for the known eruptive variable V2492 Cyg. The latter images is used as a proxy of the bright {\it Gaia} magnitude of the YSO. In the images, the location of the object is marked by the blue arrow. The images have a size of $2.5\arcmin\times1.7\arcmin$.}
    \label{fig:v2492Cyg_a}
\end{figure*}

To understand the types of variability that we would detect with this method, we analysed a subsample of 5000 YSOs\footnote{This sample is selected from some of the categories described at the beginning of Section \ref{sec:sample}. We note that these do not include the FU Orionis star class.}  as well as a sample of known variable YSOs taken from the {\it Gaia} Photometric Science Alerts\footnote{http://gsaweb.ast.cam.ac.uk/alerts} \citep[GSA][]{2012Wyrzykowski} which include a number of objects that are known eruptive YSOs. We note that in both samples the evolutionary class of the individual YSOs was not taken into account.

This analysis yielded 24 bonafide high-amplitude ($\Delta$R$\geq1.5$~mag) variable stars from the SIMBAD sub-sample and successfully retrieved the {\it Gaia} YSO alerts showing high-amplitude variability with $\Delta G> 2$~mag.

We found that most variable stars from the SIMBAD and {\it Gaia} alerts sub-samples show low-amplitude, recurrent aperiodic variability. A literature search revealed that the variability in these low-amplitude objects is caused by physical processes other than variable accretion. The low-amplitude objects showed lower values of $\Delta_{\mathrm{YSO}}$, whilst a few of these had positive values of $\Delta_{\mathrm{YSO}}$, but with individual values of $G-M_{\rm SSS}$, when analysed separately as a univariate distribution, that were found to be below $5\sigma$ in each band.

In the cases where we suspected variability was due to large accretion events, and in the known eruptive YSOs from the {\it Gaia} alerts, we found that $\Delta_{\mathrm{YSO}}$ showed a value much larger than 0. In addition the inspection of individual values of $G-M_{\rm sss}$, when analysed separately as a univariate distribution, were found to be above $5\sigma$ in at least one band. 

Given the results from this early analysis, when applying our method to the class II YSO sample, we defined a new set of conditions for an object to be included as a candidate variable star. This was done to reduce the number of objects to inspect (see below) and to try and include only the most extreme cases of variability, and thus more likely related to large and long-lasting changes in the accretion rate. The revised conditions were as follows.

\begin{enumerate}

\item $\Delta_{\mathrm{YSO}}\geq70$ and the individual values of $G-M_{\rm sss}$, when analysed separately as a univariate distribution, to be above $5\sigma$ in at least one band.

\item $\Delta_{\mathrm{YSO}}\geq100$ and the individual values of $G-M_{\rm sss}$, when analysed separately as a univariate distribution, to be below $5\sigma$ in each band.

\end{enumerate}

\subsection{Problems with the photographic images images}\label{sec:plates}

Through the initial analysis of the two YSO subsets we were able to recognise a major source of contamination in our variable star candidates list. We found that in many objects the non-detection of the source in the SSS catalogues was probably due to problems with the plate images rather than true variability of the source. These issues were most apparent in the areas with high extinction, and were more evident in the $B$ and $R$ SSS images. In order to define whether the variability was driven by this issue, we first queried {\it Gaia} DR2 and the SSA to select sources within a box size of $3\arcmin\times3\arcmin$ around the YSO being analysed. Then we crossmatched both catalogues to determine the number of objects detected both in SSS and {\it Gaia}, $\mathrm{N}_{\rm sss}$. In individual filters, YSOs found in problematic areas showed a low value for the ratio $\mathrm{N}_{\rm sss}/\mathrm{N}_{\rm Gaia}$, with $\mathrm{N}_{\rm Gaia}$ the total number of {\it Gaia} sources in the $3\arcmin\times3\arcmin$ box. Therefore, if the individual bands driving the variability of the YSO showed $\mathrm{N}_{\rm sss}/\mathrm{N}_{Gaia}<0.45$ then this object was classified as unlikely to be a variable star.

\subsection{Visual inspection of variable star candidates}\label{sec:lc_class}
We next used the results from our subsamples to inform a search for high-amplitude variable stars using all the 15404 YSOs in our sample. The method yielded 4815 objects that fulfilled the condition established by Equation \ref{eq:varcon}. This number reduced to 1576 objects when we imposed the conditions of Section \ref{sec:mt}. From the latter, 501 objects were flagged as being in the list due to the problems with photographic plates explained in Section \ref{sec:plates}. Thus, we were left with 1075 variable YSO candidates.

To confirm or discard its variability, each of the 1075 variable candidates was analysed individually. The analysis included the inspection of the $3\arcmin\times3\arcmin$ cutout images from the photographic plates provided by the SSA, as well as comparing with the more recent images (when available) provided by the Panoramic Survey Telescope and Rapid Response System \citep[Pan-STARRS,][]{2016Chambers} and the SkyMapper Southern Sky Survey \citep{2018Wolf}. In addition we also searched for additional photometry from publicly available catalogues through the VizieR catalog access tool \citep{2000Ochsenbein}. The list of the individual surveys found in Vizier that were used in our analysis is given in Appendix \ref{sec:vizier}. This final step revealed further sources of contamination that were not considered before the cleaning of the final sample. Most of these related to problems with bright objects, crowding or incorrect magnitudes in SSS catalogues. 

After inspection of the individual candidates, we found 139 true high-amplitude variable stars. In each case, the additional photometry from other surveys allowed us to classify the objects according to their light curves. We note that this is based only on a handful of epochs, therefore in many cases our classification is uncertain and is thus followed by a ? sign. This classification aims to select those objects where variability resembles that of long timescale outbursts. An example of the photometry obtained for each high amplitude variable is presented in Table \ref{tab:phot}. The full light curve data are presented as supplementary material.

\begin{table}
\centering
\caption{Example of the photometry obtained for each high-amplitude variable. In the tables we present the modified Julian date of the observations (estimated from the Julian date as MJD=JD$-$2400000.5), the $R$ or $r$-band magnitudes and the survey from where they are obtained. In the last column we present a note depending on whether the magnitudes are an upper limit or if these result from transforming between different filters (marked as approximate in the table). The full table is available online. }
\label{tab:phot}
\begin{tabular}{lccccc}
\hline
ID & MJD & Mag & Filter & Survey & note \\
\hline
V4 & 34270.404 & 21.0 &  R &         SSS(R1) &  upper limit \\
V4 & 47880.208 & 21.0 &  R &         SSS(R2) &  upper limit \\
V4 & 49253.445 & 18.5 &  R &         SSS(B) &   approximate  \\
V4 & 49654.347 & 18.5 &  R &         SSS(I) &   approximate  \\
V4 & 52953.981 & 17.46 &  r &       IPHAS &           -- \\
V4 & 56227.446 & 16.79 &  r &  Pan-STARRS &           -- \\
V4&  56537.365 & 16.49 &  R &         PTF &           -- \\
V4 & 57204.000 & 16.5 &  R &        {\it Gaia} &           approximate \\
\hline
\end{tabular}
\end{table}

%\section{Results}

The list of 139 high-amplitude variable stars selected with our method is shown in Table \ref{tab:allvar}. Column 1 shows the designation given by us. Since the objects are ordered by amplitude, the number gives an indication of their level of variability. Columns 2 and 3 mark the right ascension and declination of the object respectively. The SIMBAD identification and type are shown in columns 4 and 5, whilst column 6 gives the observed ($R$) amplitude taking into account only the SSS and {\it Gaia} epochs (after transforming the {\it Gaia} $G$, SSS $B$ and $I$ into $R$-band magnitudes). We note that these are approximate values. In column 7 we present a classification for the behaviour of the light curve of the YSO (Section \ref{sec:lc_class}). Column 8 provides a final classification of the light curve, which is obtained either by the analysis done in this work (see Section \ref{sec:sel_hamp}) or by information found in the literature (e.g. V26 is a known UXor), whilst column 9 shows the references to such classification. In column 10 we mark whether the object is a known variable star, whilst column 11 shows whether the SSS vs {\it Gaia} variability relates to the known variability of the source of if our detection relates to previously unknown variability of the source.  Finally column 12 shows the source that was used to classify the YSOs as being in the class II stage (see Section \ref{sec:ysoclass}).  

Our sample contains 66 new variable stars, which represents 48$\%$ of our sample. From the known variable stars (72 objects) we find that in 17 objects the SSS vs {\it Gaia} variability has not been reported previously. We note that upon further inspection objects V5, V12, V24, V32, V45 and V84 are found to be more likely non-YSOs. Therefore, although these are  shown in Table \ref{tab:allvar}, we did not include these objects in the following analysis of variability in the class II YSOs.

\section{The distribution of class II amplitudes}\label{sec:class2amp}

The top plot of Figure \ref{fig:amp_hist} shows the histogram of the amplitudes of our high-amplitude variable stars as determined from the SSS and {\it Gaia} epochs. We note that in the histogram we did not include the objects that were found to be likely non-YSOs. In addition we also did not include the three long-lasting outbursting YSOs that are more likely class I YSOs (see Section \ref{ssec:ysoc}).

In the histogram we can see that we do not detect objects with amplitudes below 1 mag. In addition the sample appears incomplete below 2 magnitudes.

To obtain the distribution of amplitudes for low amplitudes we used the dataset of $i$-band photometry of Cep OB3b presented in Sergison et al. (in prep). We took the objects classified as Class II on the basis of their Spitzer photometry \citep{2012Allen}, which are in Sergison et al's primary sample. Importantly, variability has played no part in the selection of these stars. We took one long observation from each of 2004, 2005, 2007 and 2013 as the closest analogue we could obtain to the sampling in this paper. For each star which had ``good'' photometry (as defined by Sergison et al) for all four datapoints we then calculated a full amplitude. 

To compare both samples we needed to estimate the fraction of stars per unit amplitude (d$f$/d$A$). However, given the observed amplitudes of our YSO variables, we considered the possibility that some objects in our class II YSO sample were too faint to have been detected in {\it Gaia}, even if they showed large amplitude outbursts. We found that 3712 YSOs had no {\it Gaia} DR2 counterparts, but this was found to be consistent with the fact that they were also faint or not detected in SSS, so they were flagged as non-variable stars. 

There is no detailed selection function for {\it Gaia}, but the star counts turn over at about $G\simeq20.5$~mag \citep{2018Gaiadr2}. Hence, objects with $G\leq22.5$~mag showing variability with $\Delta R\geq$2 mag should have been detected in {\it Gaia} DR2. To estimate an approximate $G$ magnitude for the 3712 YSOs that are not in {\it Gaia} DR2, we obtained sloan-$r$, $J$ and $K$ photometry of the YSOs by crossmatching with catalogues from Pan-STARRS, SkyMapper, the SDSS photometric catalogue \citep{c2015Alam},  DENIS \citep{1994Epchtein}, 2MASS \citet{c2003Cutri}, UKIDSS GPS \citep{c2008Lucas} and the VISTA Variable in the Via Lactea Survey \citep{c2017Minniti}. Objects that were detected with $r<24$~mag would have been detected in {\it Gaia} if they had shown high-amplitude variability. Objects without $r$-band detections, do have $J$, and $K$ photometry, so we use the information from YSOs that were detected in {\it Gaia} DR2 to determine a relation between the $G-J$ and $J-K$ colours. The latter fit was then used to estimate an approximate value of $G$ for the objects that are not originally detected in {\it Gaia} DR2.

This method showed that 1318 YSOs had approximate $G$-band magnitudes that were fainter than 22.5 mag, so we believe these would have not been detected even if they had shown high-amplitude variability. Thus for the remainder of our analysis of variability, our total sample of class II YSOs reduces to 14086 objects.

The comparison  of d$f$/d$A$ between our sample and that of Sergison et al. (see Figure \ref{fig:amp_hist2}) confirms that our sample is not complete below amplitudes of 1 magnitude. Both samples agree remarkably well at 2 magnitudes. The figure also shows that at $\Delta m>3.5$~mag the number of objects does not fall as expected from the rate of decline observed at lower amplitudes. Instead, the distribution reaches a plateaux, which could point to YSOs at these amplitudes being part of a different population of variable stars.

\begin{table*}
	\centering
	\caption{The 139 high-amplitude variable stars detected in our analysis. The full table is available online.}
	\label{tab:allvar}
\resizebox{\textwidth}{!}{
   	\begin{tabular}{lccccccccccc} % four columns, alignment for each
		\hline
		ID & $\alpha$ & $\delta$ & SIMBAD ID & SIMBAD class & $\Delta R$ & LC class & Final class & Reference & Known? & SSS? &YSO class  \\
		\hline
		V1 & 20:58:53.72 & $+$44:15:28.3 & EM* LkHA  190 & FUOr & 5.2 & Outburst & Long-lasting (class I) & This work (Section 5) & Y & Y & Marton \\
		V2 & 20:58:17.03 & $+$43:53:43.3 & V* V2493 Cyg & FUOr & 5.0 & Outburst &  Long-lasting (class II)  & This work (Section 5) & Y & Y & Marton \\
		V3 & 20:54:07.39 & $+$41:34:58.1 & V* V1219 Cyg & Orion V* & 4.7 & Repetitive fadings &  Deep fades  & \citet{2013Stecklum} & Y & Y & Marton \\
		V4 & 02:33:53.40 & $+$61:56:50.1 & 2MASS J02335340$+$6156501 & YSO & 4.5 & Outburst & Long-lasting (class II) & This work (Section 5)  & N & N & 2M\_AAA \\
		V5$^{\dagger}$ & 05:28:54.06 & $-$06:06:06.3 & V* RX Ori & Orion V* & 4.5 & Periodic & Non-YSO &  \citet{2007Lee}  & Y & Y & WISE\_slt04 \\
		V6 & 21:43:00.01 & $+$66:11:27.9 & V* V350 Cep & Orion V* & 4.5 & Outburst & Long-lasting (class I) & This work (Section 5)    & Y & Y & Marton \\
		V7 & 16:11:46.00 & $-$25:32:00.8 & V* V931 Sco & Orion V* & 4.4 & Irregular & Periodic (P=215 d) & \citet{2002Pojmanski} & Y & Y & Marton \\
		V8 & 05:10:11.00 & $-$03:28:26.2 & 2MASS J05101100$-$0328262 & YSO & 4.0 & Irregular & Short-term outburst & \citet{2017Sicilia}  & Y & Y & WISE\_slt04 \\
		V9 & 08:41:06.76 & $-$40:52:17.4 & 2MASS J08410676$-$4052174 & Em* & 4.0 & Outburst &  Long-lasting (class II) & This work (Section 5)    & Y & N & Marton \\
		V10 & 22:53:33.26 & $+$62:32:23.6 & V* V733 Cep & FUOr & 4.0 & Outburst &  Long-lasting (class II) & This work (Section 5)   & Y & Y & Marton \\
		V11 & 05:37:00.10 & $-$06:33:27.3 & V* BE Ori & Em* & 4.0 & Repetitive outbursts? &  --   &  -- & Y & Y & Megeath\_alpha \\
		V12$^{\dagger}$ & 16:00:07.42 & $-$41:49:48.4 & 2MASS J16000742$-$4149484 & Candidate YSO & 3.8 & Irregular &  Non-YSO & \citet{2017Frasca}   & N & N & C2D \\
		\hline
		\multicolumn{11}{l}{$\dagger$ Objects not included in any subsequent analysis.}\\
	\end{tabular}}
\end{table*}

\section{Selecting high amplitude accretion-driven variability}\label{sec:sel_hamp}

We detected a large number of high-amplitude variable YSOs. The next step consisted of determining what is the physical mechanism driving these large changes.

\subsection{Physical mechanism}

Variability is one of the defining characteristics of pre-main-sequence stars \citep[e.g. ][]{1945Joy}. This occurs on a broad range of timescales and amplitudes due to various physical mechanisms. Periodic modulation of the stellar flux resulting from the rotation of cool spots in the stellar photosphere is mostly observed in weakly accreting YSOs (the so-called weak-lined T Tauri stars or WTTS) and its characterised by low level variability in the order of 0.1 mag \citep{1994Herbst}. The irregular (although sometimes periodic) variability from short-lived hot spots at the stellar surface due to accretion can reach can 2-3 mags in extreme cases \citep[see e.g. ][]{2007Grankin}. Luminosity dips caused by circumstellar dust obscuration can last days to months and with amplitudes that are effectively limitless as they depend of the optical depth of the dust obscuring the central star \citep[see e.g ][]{2001Carpenter}. Finally, changes in the accretion rate from the disc onto the central star can also cause variability, with timescales of days to hundreds of years, in YSOs (see Section \ref{sec:intro}).

% \citet{1994Herbst} defines three types of variability in days to weeks timescales in these objects: periodic modulation of the stellar flux resulting from the rotation of cool spots in the stellar photosphere, irregular (although sometimes periodic) variability from short-lived hot spots at the stellar surface due to accretion and luminosity dips caused circumstellar dust obscuration. The first group is mostly observed in weakly accreting YSOs (the so-called weak-lined T Tauri stars or WTTS) and its characterised by low level variability in the order of a few 0.1 mag. The amplitudes of the variability due to hot spots can reach 2-3 mags in extreme case of objects in the second group \citep[see e.g. ][]{2007Grankin}, whilst the third group shows luminosity dips that can last days to months and with amplitudes that are effectively limitless as they depend of the optical depth of the dust obscuring the central star \citep[see e.g ][]{2001Carpenter}.  Finally, changes in the accretion rate from the disk onto the central star can also cause variability , with timescales of days to hundreds of years, in YSOs (see Section \ref{sec:intro}).
%Inspection of the light curves of the variable 

The amplitude of the variability of stars in our sample gives us an insight into the physical mechanism that drives the variability in these YSOs. We have previously discussed that at amplitudes below 2 magnitudes the sample is incomplete. However, this is not a problem for the purposes of this study as we are interested in the most extreme cases of accretion-related variability which we expect to have amplitudes greater than this level. We would not expect to detect YSOs where variability is caused by cool spots in the stellar photosphere, as amplitudes due to this effect are not expected to reach more than a few tenths of a magnitude. In addition our sample is comprised of class II YSOs, where this effect is not expected to dominate. The lower number of YSOs at amplitudes below 2 magnitudes and the non-detection at below 1 mag, allows us to conclude that we are not observing variability due to cool spots. 

\begin{figure}
	\resizebox{0.95\columnwidth}{!}{\includegraphics{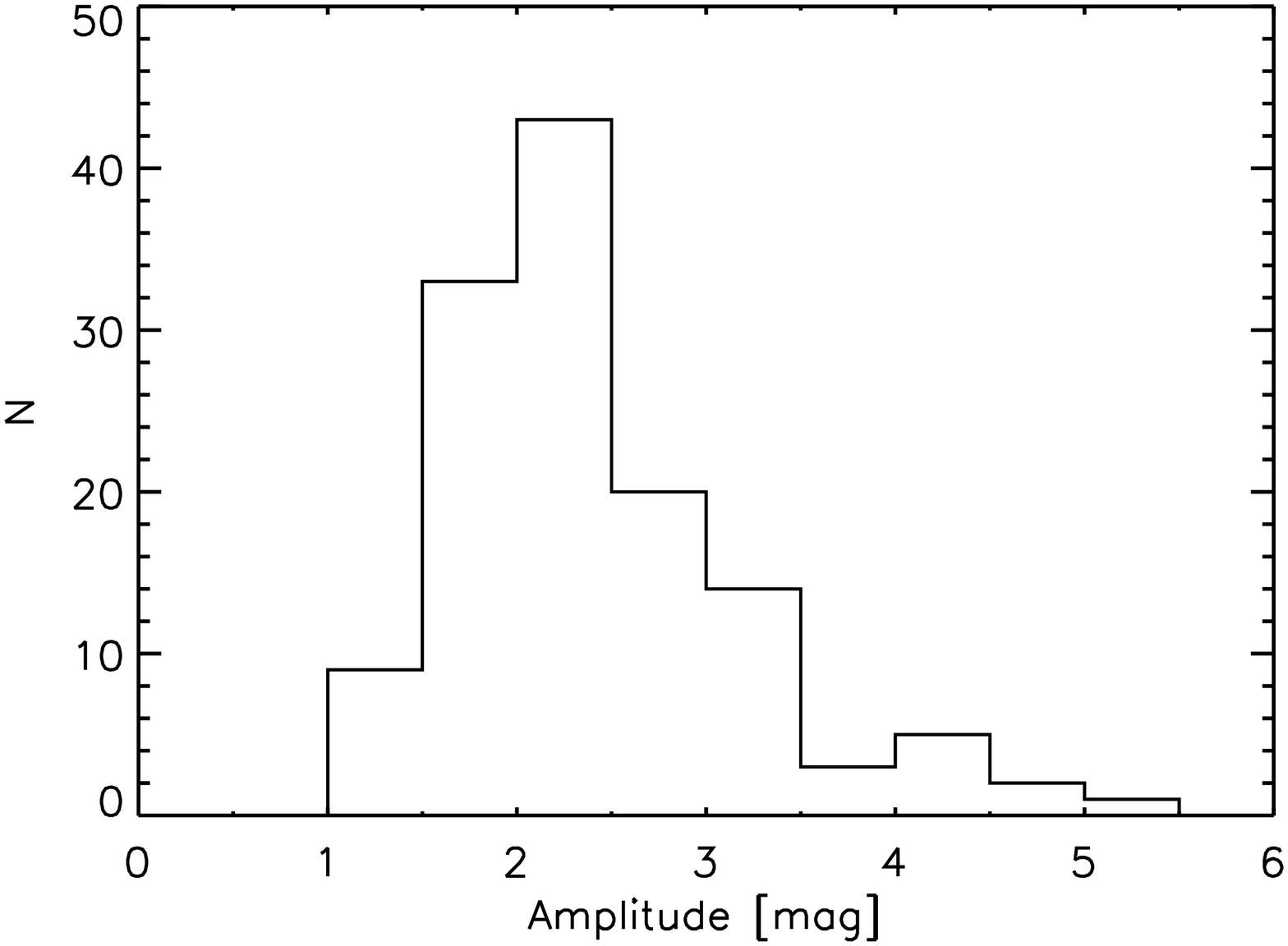}}\\
	\resizebox{0.95\columnwidth}{!}{\includegraphics{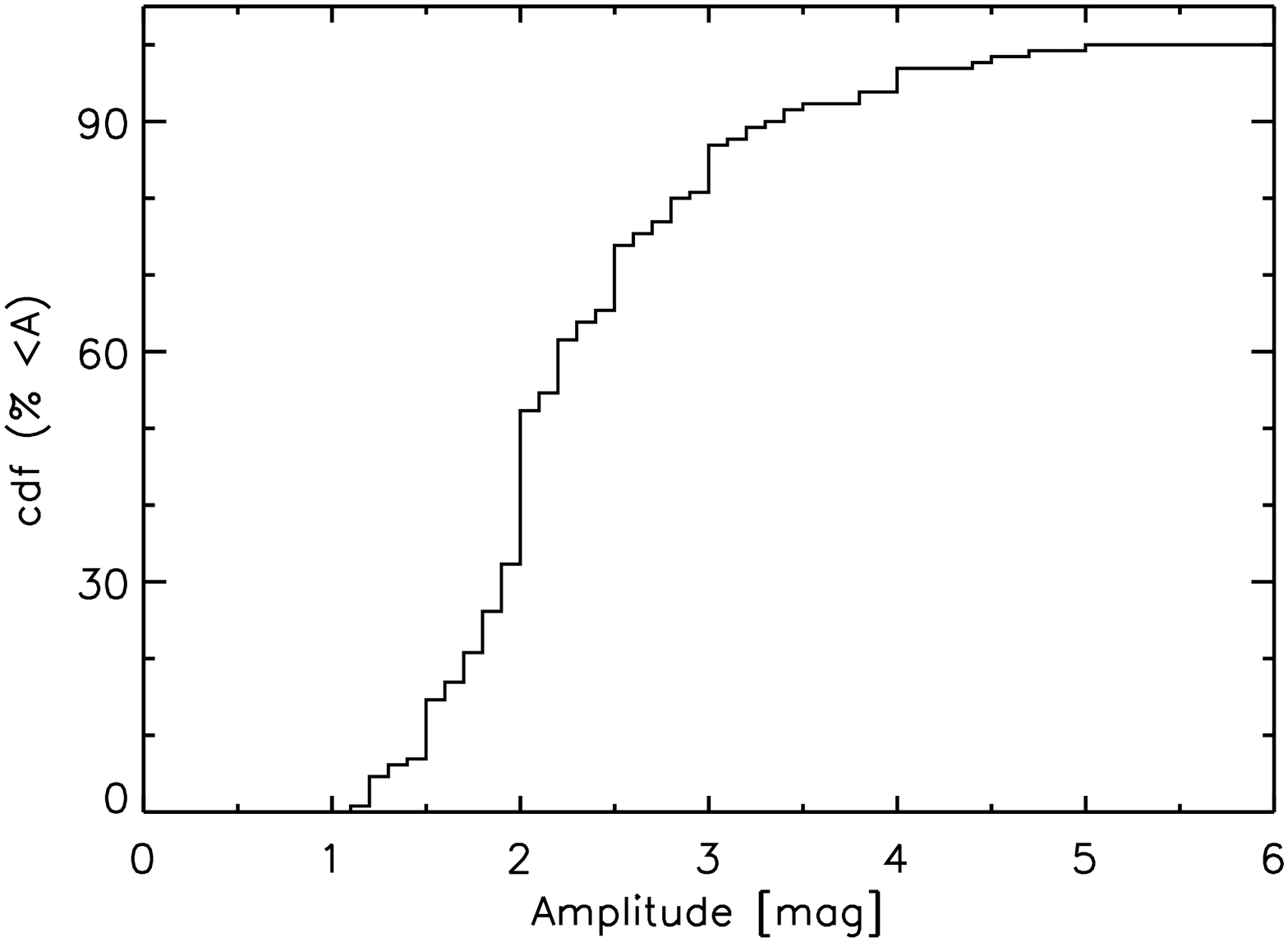}}
    \caption{(top) $R$-band amplitude histogram for the high-amplitude variable stars recovered by our search. (bottom) Cumulative distribution of the $R$-band amplitude, representing the percentage of objects in our sample that are found  below the corresponding magnitude level.}
    \label{fig:amp_hist}
\end{figure}

\begin{figure*}
\resizebox{0.7\textwidth}{!}{\includegraphics{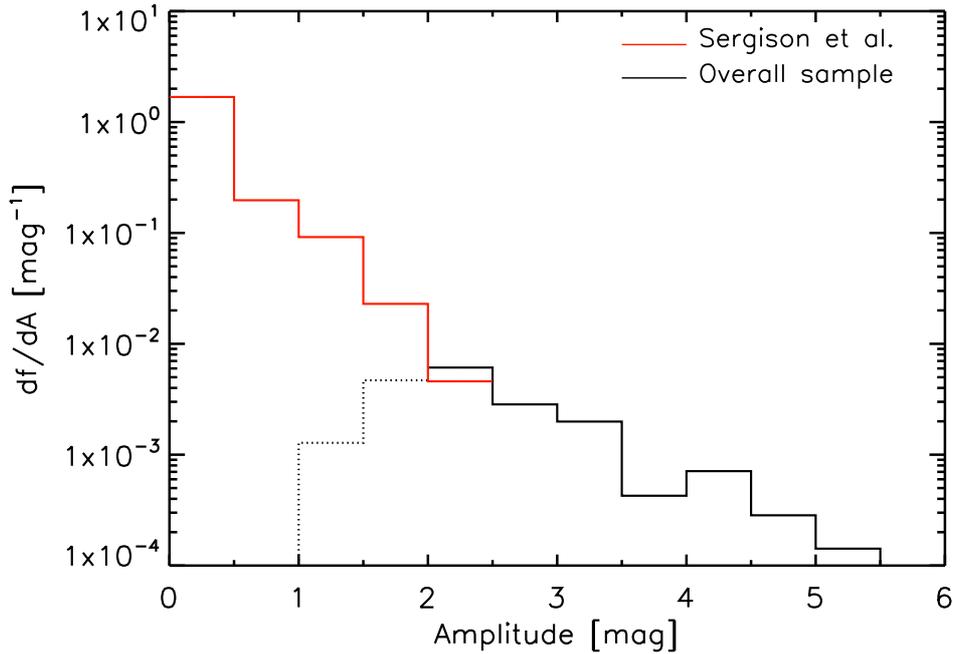}}
\caption{Fraction of stars per unit amplitude, d$f$/d$A$. The solid red line shows the distribution for variable YSOs from Cep OB3b presented in Sergison et al (in prep), whilst the solid black line marks the distribution for YSOs in this work. The dotted black line also marks the distribution of variable YSOs from this work, but for objects in the region where we are incomplete (see main text).}
    \label{fig:amp_hist2}
\end{figure*}

Figures \ref{fig:amp_hist} and \ref{fig:amp_hist2} also show that the number of variable YSOs drops dramatically at amplitudes larger than 3.5 mag, but there appears to be a secondary peak at around 4.5 mag, although the position of the peak is highly dependent on the choice of bin size. However the cumulative distribution shown in Figure \ref{fig:amp_hist} does reveal a bump at around 4 mag. Thus the variable YSOs with amplitudes larger than 3.5 magnitudes could be a different population from the objects described above. Considering the photometric characteristics of known long-lasting YSO outbursts, it is at these amplitudes that we would expect to find the majority of objects that can be classified as showing long-term, accretion-driven outbursts. 

\subsection{Selecting long timescale variability}\label{ssec:longvar}

The selection of objects where variability is driven by extreme, long-term changes in the accretion rate was difficult. In many cases using just the data from SSS  and {\it Gaia}  (5 epochs of observations) could give a false impression that the object is an eruptive YSO. Therefore, we made use  of all the available photometry (see Section \ref{sec:lc_class})  to discard or confirm a classification as long-lasting YSO outbursts. In many cases the classification was aided by a more exhaustive literature search.

In this analysis we found that the majority of YSOs with amplitudes between 1 and 3 mag (101 out of 108 objects) are characterised by light curves with fading events or irregular variability (i.e. going from bright to faint states repetitively across their light curves). The variability in this group is likely to be explained by hot spots due to accretion, representing extreme cases of this type of variability \citep[see][]{2007Grankin} or from dust obscuration, similar to the fading events observed in AA Tau \citep{2013Bouvier} or  in RW Aur \citep[e.g. ][]{2016Bozhinova}.  The remaining 7 YSOs (V36, V51, V53, V56, V68, V85 and V94) were classified as a potentially having long lasting outbursts based on their light curves.

From the 28 objects with amplitudes of $\Delta R\geq3$~mag (variable stars V1 to V4, V6 to V11, V13 to V23, and V25 to V31), 7 were classified as showing long-lasting outbursts. Amongst the 28 objects we also find objects with apparent long-lasting fading events, again likely related to extinction, and objects with irregular variability. However, given the larger amplitudes, variability due to hot-spots is less likely to explain this kind of variability. Instead these are more likely explained by short-term accretion events that appear as irregular given the long baseline of the light curves.

\begin{figure*}
	\resizebox{0.9\textwidth}{!}{\includegraphics{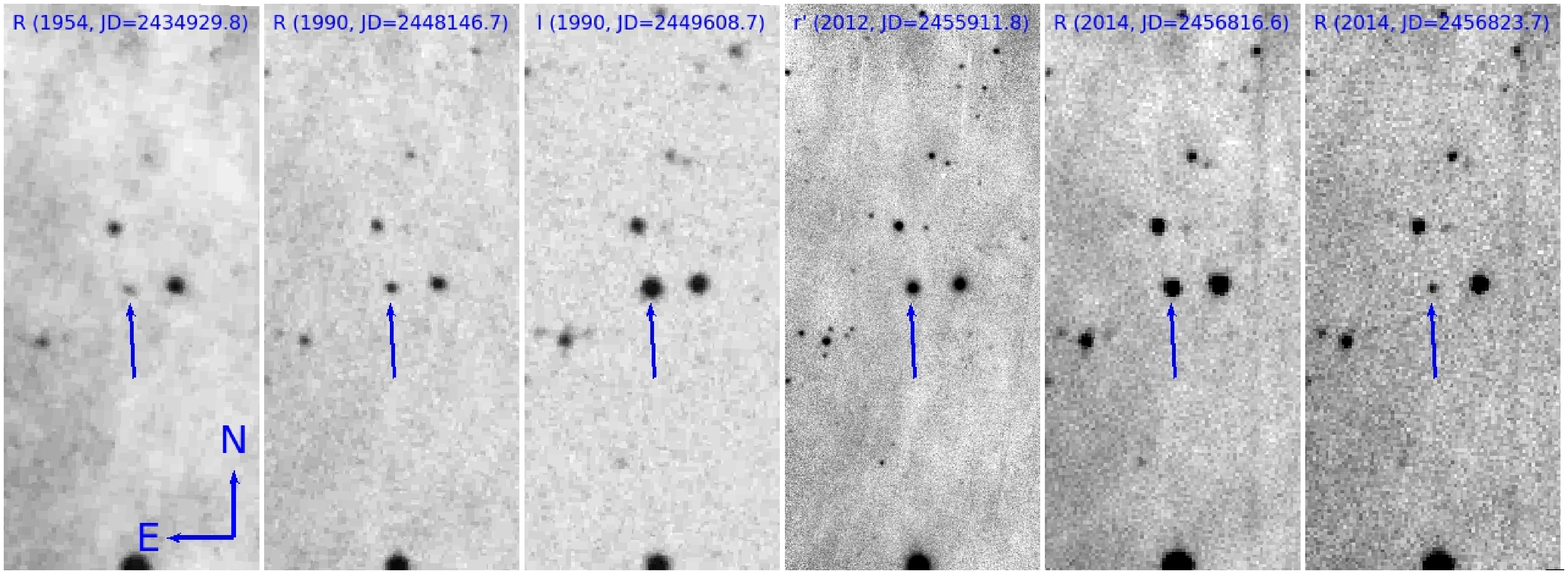}}\\
	\resizebox{0.8\textwidth}{!}{\includegraphics{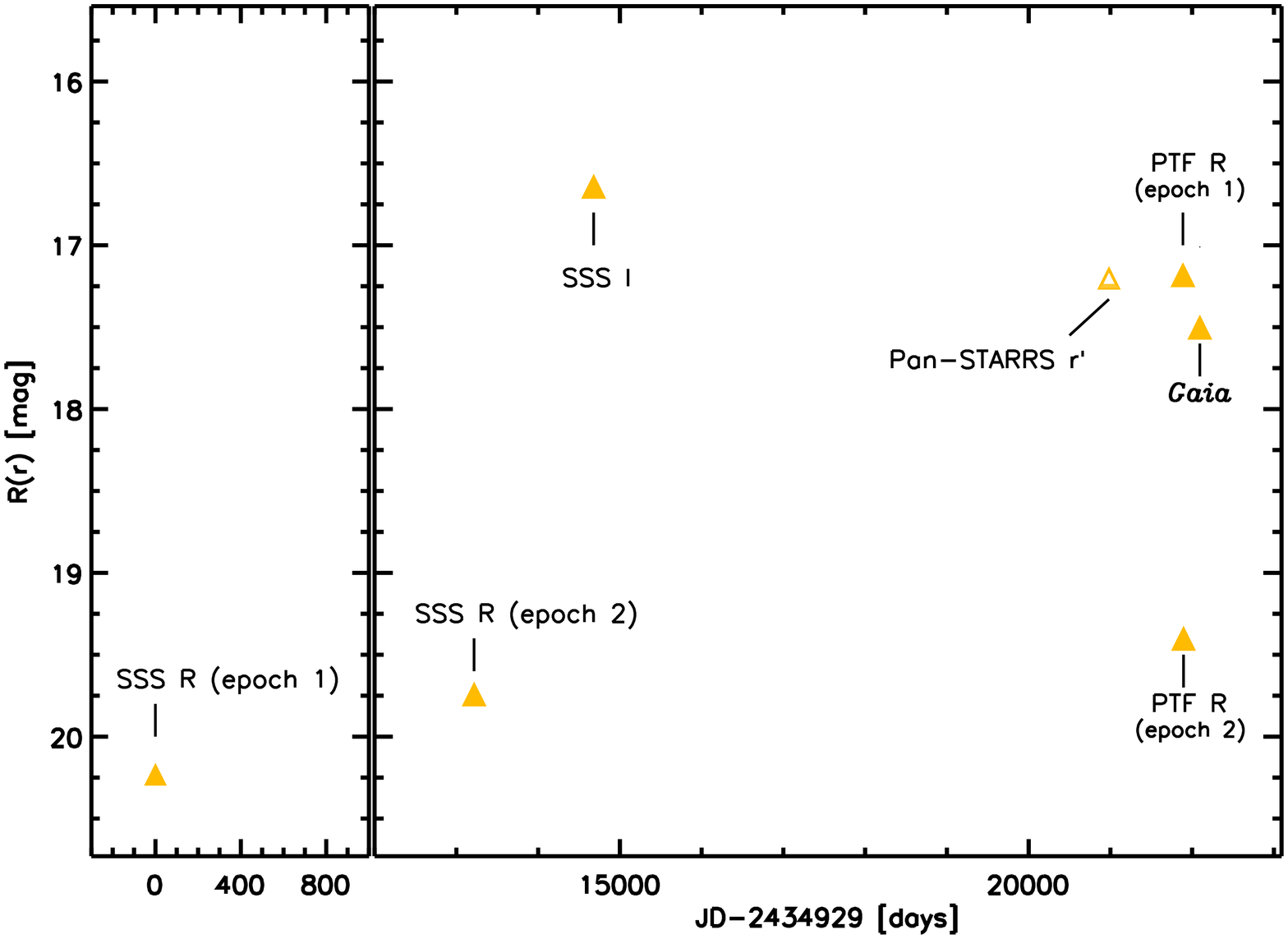}}
	%\resizebox{0.6\textwidth}{!}{\includegraphics{v43images_b.eps}}
    \caption{(top) Comparison of POSS-I $R$, POSS-II $R$ and POSS-II $I$ photographic plate images with sloan $r$ image from Pan-STARRS (as a proxy for the {\it Gaia} magnitude) and $R$-band images from the PTF survey. We mark the date of observations at the top of each image, whilst the variable star is marked by a blue arrow. All of the images have a size of $3\arcmin\times1.3\arcmin $. (bottom) Light curve of V20 showing the $R$ (filled triangles) and sloan $r$ (open triangle) magnitudes. The photometry from the SSS $I$ and {\it Gaia} $G$ filters have been converted to approximate R-band magnitudes to help the clarity of the light curve. In the plot we mark the approximate epoch of the photometry for which we present images in the top plot of the figure.}
    \label{fig:v43}
\end{figure*}

The variable star V20 (2MASS J20521294$+$4420534) illustrates the issues of classifying YSOs based only a handful of epochs. Figure \ref{fig:v43} shows the light curve of the source using the SSS and {\it Gaia} epochs, in addition we also compare the $R$ and $I$ images from SSS. Based only on this information, this object appears to have gone into outburst after the second epoch of $R$-band observations, becoming $\simeq$3.2 mag brighter during the I SSS observations and remaining bright until the {\it Gaia} observations. However, the addition of literature data and inspection of images provided by the Palomar Transient Factory \citep[PTF][]{2009Law} reveals that the faint $R$ magnitudes are probably part of the ``normal'' variability of the source, which seems to be characterised by irregular, short-term variability (see the bottom images of Figure \ref{fig:v43}).

\begin{figure}
	\resizebox{0.95\columnwidth}{!}{\includegraphics{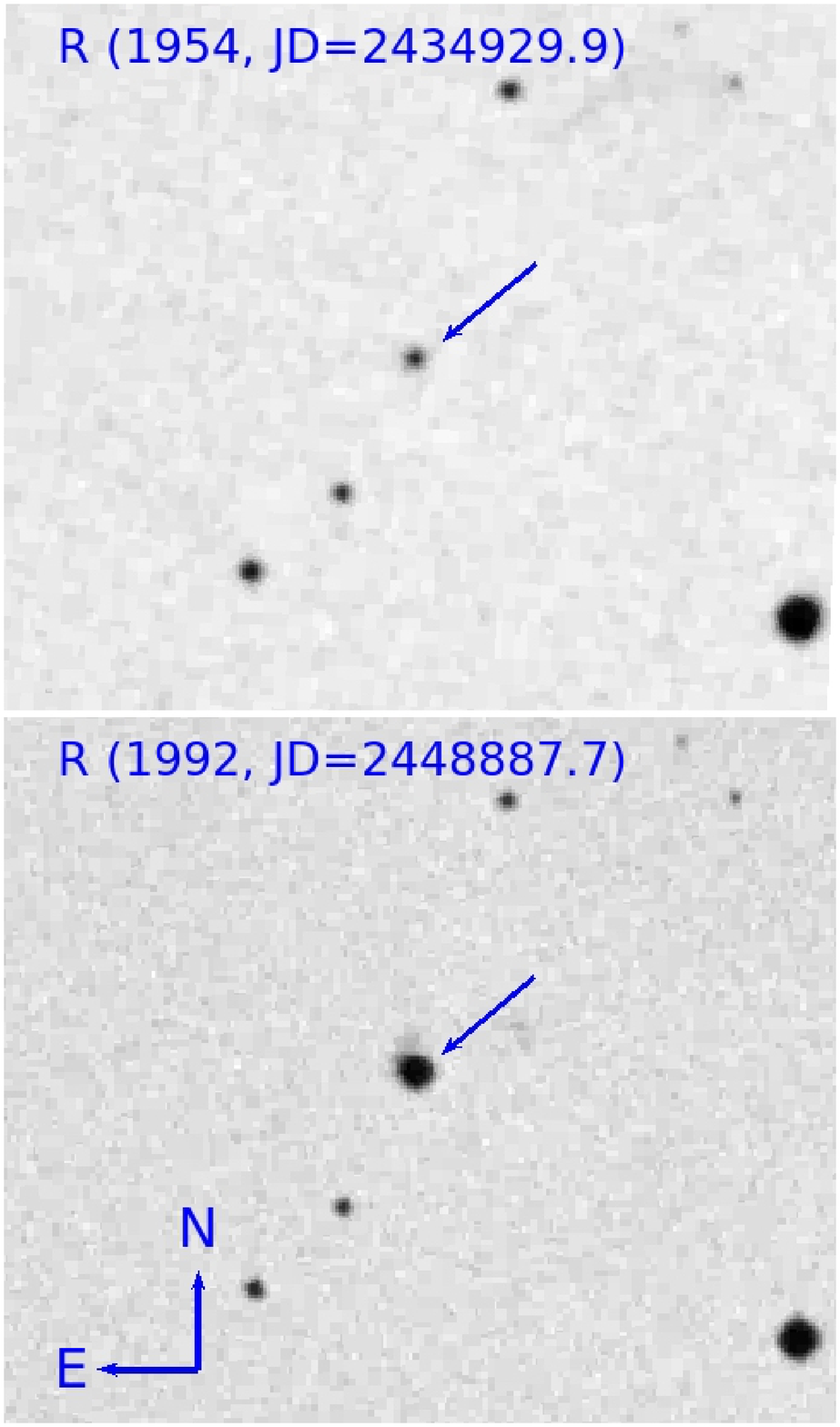}}
    \caption{Comparison of the POSS-I E and POSS-II R $3\arcmin\times3\arcmin$ images of V3, obtained in 1954 and 1992 respectively. In both images the variable star is marked by the blue arrow. }
    \label{fig:v18}
\end{figure}

A second example is V3 (V1219 Cyg), which could be easily mistaken as showing a 4.7 mag outburst between the first and second epochs of $R$ SSS observations (see Figure \ref{fig:v18}), in addition the object drives an H$_{2}$ outflow \citep{2018Makin}. However, further analysis reveals that this source has been classified in the past as showing repetitive high-amplitude fading events rather than a long timescale outburst \citep[see][]{2013Stecklum}.

In both examples the light curves, given only the SSS and {\it Gaia} data, could easily be classified as long-term outbursts. We were able to discard such classification only through the inclusion of additional data from the literature. 

Therefore in trying to find objects with long-lasting outbursts, we discovered a large population of high-amplitude variable YSOs that cannot be classified as showing long-term outbursts (see Section \ref{ssec:outb}). Changes in the accretion rate could still be the underlying cause of such variability, but objects that show repetitive and/or short-term (lasting less than $\simeq $10 yr) outbursts, such as V18 or V8 \citep[the known short-term eruptive YSO ASASSN13-db, see][]{2017Sicilia}, are not included in the subsequent analysis. However, we emphasise that these are an interesting class of objects in their own right.

In addition, many objects show apparent long-lasting fading events. This type of variability is explained by obscuration of the central star by structures in the disc located at large distances and above the midplane of the disc \citep[see e.g.][]{2013Bouvier}. This is interesting in the context of discs and planet formation. The migration of giant planets due to interactions with the disc \citep[also known as type II migration,][]{2004Ida} can open up gaps in the disc which could lead to scaleheight bumps \citep{2013Baruteau} or to a misalignment between the inner and outer disc \citep{2016Lubow}, that could lead to fading events. We note, however, that mechanisms such as dusty outflows could also lead to these type of events \citep{2018Davies}.

Again, these objects are interesting findings, but further analysis of them is beyond the scope of this paper.

\subsection{Outbursts}\label{ssec:outb}

To determine an outburst rate from our analysis, we needed to select only objects where a classification as a long-lasting YSO outburst was very likely. In this sense YSOs where changes in the accretion rate might be the physical mechanism driving the variability, but which show repetitive and/or short-term outbursts are not included (see above). Additionally we did not include objects where lack of data did not allow a firm classification. For example, five out of seven objects with $\Delta R \leq 3$ mag and that were classified as potentially having long-lasting outbursts, were not included in our final sample as there was not enough data that would allow to confirm such classification.

Collating literature and database information in this way allowed us to classify 9 objects as being part of the eruptive YSO class showing long-lasting outbursts (the 7 objects with $\Delta R\geq3$~mag classified as having long-lasting outbursts in Section \ref{ssec:longvar} and two objects with $\Delta R<3$~mag, V36 and V51). These include 6 known eruptive YSOs, V1 \citep[LkHA 190, most commonly known as V1057 Cyg, e.g.][]{1987Hartmann},  V2 \citep[V2493 Cyg, most commonly known as HBC722,][]{2010Semkov}, V6 \citep[V350 Cep,][]{2014Ibryamov}, V10 \citep[V733 Cep,][]{2007Reipurth}, V31 \citep[UCAC2  30330609, most commonly known as V960 Mon,][]{2014Hillenbrand} and  V36 \citep[V582 Aur,][]{2018Abraham}. There are 3 previously unknown eruptive YSOs, V4 (2MASS J02335340$+$6156501), V9 (2MASS J08410676$-$4052174) and V51 (WRAY 15-488). A more extensive explanation for the classification of the previously unknown objects is presented as individual notes in Appendix \ref{app:a}.

\subsection{Class I contamination}\label{ssec:ysoc}  

Due to the effects of geometry, reddening or disc inclination there might not be a direct correlation between the YSO class and its evolutionary stage \citep[see e.g.][]{2006Robitaille}. The classification of objects in Section \ref{sec:ysoclass} was done using near- to mid-infrared data, which lead to a classification as class II YSOs, however, longer wavelengths could reveal the presence of envelopes (see also Section \ref{ssec:ysomisc} for further discussions on the contamination from class I YSOs).

Therefore, before we determined the outburst rate from our selected YSOs, we took into consideration the possible contamination from YSOs at earlier evolutionary stages in our sample (class I or flat-spectrum sources). The concern arises from the fact that if the outbursts are $\simeq$10 times more common during the class I stage (see Section \ref{sssec:scholz13}), then even a low contamination from class I YSOs would result in a high fraction of our outbursts sources being from the embedded phase.

To study this effect we performed a deeper analysis of the likely evolutionary stage of the YSOs in our long-term outburst sample. From the 9 YSOs in this sample, 8 were selected from the \citet{2016Marton} catalogue and given a class II YSO classification using the \citet{2014Koenig} criteria, whilst V4 is selected from the near-infrared excess as determined from its 2MASS colours. We determined the 2--24 $\mu$m spectral index, $\alpha$, for the observed  spectral energy distribution of the burst sample, using 2MASS and {\it WISE} magnitudes. Table \ref{tab:outb_alpha} lists the values determined for the sample. We see that $\alpha\leq-0.3$ for four objects in the sample is consistent with the definition for class II YSOs of \citet{1994Greene} . For five objects, the spectral index indicates an earlier evolutionary stage. 

For objects where $\alpha$ is consistent with a class II stage, we find further evidence to support the idea that the envelope has dissipated and that these are pure disc-bearing objects. The interferometric $^{13}$CO and C$^{18}$O survey of \citet{2017Feher} determines that V10 (V733 Cep) and V2 (HBC722) have evolved beyond the class I and flat-spectrum stages. \citet{2018Abraham} determines that the pre-outburst spectrum of V36 (V582 Aur) is that of a classical T Tauri star. There is no literature information in the case of V4, however the spectral index of $\alpha=-0.4$ is an upper limit and makes this object a very likely class II YSO.

\begin{table}
\centering
\caption{Final sample of YSO outbursts.}
\label{tab:outb_alpha}
\begin{tabular}{llcl} % four columns, alignment for each
\hline
ID  & Name$^{\dagger}$ & $\alpha$ & Envelope? \\
\hline
V1 & V1057 Cyg & $-0.1$ & Yes \citep{2017Feher} \\
V2  & HBC 722 &$-0.44$ &No \citep{2017Feher} \\
V4 & 2M J0233$+$6156 & $-0.4$ & ? \\
V6  & V350 Cep & $-0.17$ & Yes \citep{2004Muzerolle}  \\
V9  & 2M J0841$-$4052 &$-0.23$ & ? \\
V10  &  V733 Cep & $-0.78$ & No \citep{2017Feher} \\
V31  & V960 Mon & $-0.19$ & Yes \citep{2015Kospal}\\
V36  & V582 Aur & $-0.33$ & No \citep{2018Abraham}\\
V51 & WRAY 15-488 & $-0.27$ & ? \\
\hline
\multicolumn{4}{l}{${\dagger}$ We do not present the SIMBAD ID of the variables stars from}\\
\multicolumn{4}{l}{Table \ref{tab:allvar}, but instead the column shows the most common name}\\
\multicolumn{4}{l}{found in the literature.} 
\end{tabular}
\end{table}

%SSS\_v1 &  \citeauthor{2016Marton}$+$\citeauthor{2014Koenig} & $-0.1$ & Yes \\

In the case of objects with $\alpha>-0.3$  we find that V1 (V1057 Cyg) shows evidence for a large and dense envelope \citep{2017Feher}, whilst the SED of V31 (V960 Mon) suggests the presence of an envelope in \citet{2015Kospal}. V6 (V350 Cep) is classified as a protostar candidate in \citet{2004Muzerolle} (based also on the 70$\mu$m magnitude of the YSO) and \citet{1994Miranda} establish that the object is embedded in nebulosity. Thus, these 3 YSOs were not included in our calculation of the outburst rate (see Section \ref{sec:outrate}).
 
The value of spectral index for V9 and V51 is not within the class II limits of the \citet{1994Greene} classification. However, the value is not far from those of class II YSOs, and we note that we did not consider the effects of reddening in the calculation of $\alpha$, which would likely place the objects in the class II definition. In addition is hard to determine whether the slope of the SED has changed due to the outburst. In the case of V9 we find that  from observations prior to outburst, the object is an H$\alpha$ emission line star from \citet{1994Pettersson}. The authors provide a measure of the strength of the H$\alpha$ line on a scale from 1 (faint) to 5 (very strong), where V9 was listed with I(H$\alpha)=3$. We can estimate the equivalent width (EW) this corresponds to using objects from \citet{1994Pettersson} with measured EWs which also have I(H$\alpha$) classes (see their table 3). YSOs that show the same value for the intensity of the emission line, have EWs that go between 10 and 100\AA~ with an average of 38\AA. Thus, is very likely that the progenitor of V9 is in fact a disc-bearing (class II) YSO. Therefore the object was retained in the outburst sample for the remainder of our analysis. V51 was also originally classified as an emission line star prior to outburst \citep{1966Wray}. In addition the object was classified as a classical T Tauri star (from it SED during outburst, see Appendix \ref{app:av51}) by \citet{2002Gregorio}.  Therefore we included both objects in our sample of class II outbursts. Thus we have a final sample of 6 large-amplitude, long-timescale accretion driven class II YSO outbursts.

\section{Outburst Rate}\label{sec:outrate}

To determine the timescale of high-amplitude, long-lasting outbursts events during the class II stage, we first obtained the baseline of observation for each YSO in our sample. Therefore, we queried the SSA database for the epoch of observation for the corresponding B, R1, R2 and I photographic plates. The baseline of observations for each star was given by the difference between the {\it Gaia} DR2 epoch (set at 2015.5) and that of the oldest photographic plate observation. The total time covered was calculated as the simple sum of all of the baselines, which yields 782959 years or a mean baseline of 55.6 years. 

We can place formal limits on the ``waiting time" ($\tau$) between outbursts, provided that our model is that all YSOs have quiescent periods of order centuries or longer, punctuated by outbursts where the star rises by more than four magnitudes on a timescale of around a decade or less, and then remains bright for decades.
If that is the case, then the probability of observing $k$ rises, given an event rate for the whole sample of $R$ yr$^{-1}$, and a total observation time $t$ is given by the Poisson distribution
\begin{equation}
P(k|R) = {{(Rt)^k}\over{k!}}e^{-Rt}.
\label{eqn:poisson}
\end{equation}
This is formally derived by considering the binomial distribution and approximating for low event rates.
We are interested not in the event rate for the whole sample, but in the waiting time for a single star $\tau$=$1/(RN)$.
Substituting this into Equation \ref{eqn:poisson}, we then used Bayes' theorem to derive the probability density function $p(\tau|k)$ of a particular waiting time $\tau$, given $k$ as
\begin{equation}
p(\tau|k) ={{p(\tau)P(k|\tau)}\over{P(k)}} = {p(\tau)\over{P(k)}}  {{{(Nt/\tau)^k}\over{k!}}{e^{-Nt/\tau}}}.
\end{equation}
We then evaluated $P(k)$ in the normal way as
\begin{equation}
P(k)=\int_{\tau=0}^\infty {{{(Nt/\tau)^ke^{-Nt/\tau}}}\over{k!}}p(\tau){\rm d}\tau = {{Nt}\over{k(k-1)}}p(\tau),
\end{equation}
where we assumed that the prior is independent of $\tau$ (an assumption we shall return to later), and hence solved the integral using the $\Gamma$-function.
This yielded the result that
\begin{equation}
p(\tau|k) = {{{(Nt)^{k-1}}\over{(k-2)!\,\tau^k}}{e^{-Nt/\tau}}}.
\label{eqn:pdf}
\end{equation}
Differentiating this to find the turning point gave the unsurprising result that the most probable value of $\tau$ is given by
\begin{equation}
\tau={{Nt}\over k}.
\end{equation}
We then calculated the confidence limits as the values of $\tau$ which enclosed the most likely 68 percent of the probability given by Equation \ref{eqn:pdf}.
We achieved this by integrating Equation \ref{eqn:pdf} over all values of $\tau$ where $p$ exceeded a certain value.
We then decreased that value of $p$ until the integral reached 0.68, at which point the extreme values of $\tau$ are our 68 precent confidence limits.
Our observed values of $N$=14\,077, $t$=55.6 years and $k$=6 gave $\tau$=$131^{+91}_{-48}$\,kyr (68 percent confidence).

To test the effect of priors, we repeated the calculation using $p(r)\propto 1/\tau$, which gave 
\begin{equation}
\label{eqn:or8}
p(\tau | k) ={{(Nt)^k}\over{(k-1)!\,\tau^{k+1}}}{e^{-Nt/\tau}},
\end{equation}
and the most probable value of $\tau$ as
\begin{equation}
\tau={{Nt}\over (k+1)}.
\end{equation}
These yielded $\tau$=$112^{+68}_{-38}$\,kyr (68 percent confidence).
The prior is equivalent to assuming that equal decades of $\tau$ have equal probabilities of containing $\tau$.
This is normally thought a more reasonable prior for a quantity which must be greater than zero, and so this is our preferred result, although the prior does not have a great impact.

%The 9 outbursting YSOs then implies an outburst recurrence timescale of $\simeq86000$ years during the class II phase.

\subsection{Comparison with literature data}

The number we obtain in the above analysis represents the first observational determination that long-lasting outbursts do in fact occur during the class II stage and the frequency of such events. To our knowledge, there is only one other observational study, from \citet{2013Scholz}, that provides an outburst recurrence timescale for YSOs. However, our analysis differs from \citet{2013Scholz} as the authors do not divide their YSO sample into different classes.

In this subsection we compare our results with that of \citet{2013Scholz} as well as comparing it with other observational and theoretical evidence that give some estimate of the frequency of YSO outbursts.

\subsubsection{\citet{2013Scholz}}\label{sssec:scholz13} 

With the aim of determining the outburst recurrence timescale in YSOs \citet{2013Scholz} searched for young eruptive variables by comparing {\it Spitzer} with {\it WISE} photometry, thus providing a 5 year baseline. The authors performed their analysis for about 8000 YSOs divided into two samples that were analysed separately. Sample A contains 4000 YSOs selected from c2d and clusters  surveys \citep{2009Gutermuth}, with additional objects found in the NGC2264, Taurus and North American/Pelican Nebulae regions. Sample B comprises YSOs found in the \citet{2008Robitaille} catalogue of intrinsically red objects, in which \citeauthor{2013Scholz} find $\simeq 4000$ YSOs. 

In their analysis \cite{2013Scholz} did not divide objects into different YSO classes, but we require such a division to compare their data with our results.  According to the authors, sample A contains about 2800 class II YSOs. In this sample they detected one very likely outburst candidate, V2492 Cyg, which is a class I YSO \citep{2011Kospal}. 
Given the lack of class II outbursts, we cannot use the same arguments from Section \ref{sec:outrate} as the integral of Equation \ref{eqn:or8} over $\tau>0$ is not finite for k=0. All we can say is that the outburst rate is probably longer than the monitored YSO years from sample A, i.e. $5\times2800=14$~kyr.  From sample A, we can only estimate the outburst rate during the class I stage. To obtain this we have used the same $1/\tau$ prior we preferred in Section \ref{sec:outrate}, but also note that without this prior we cannot obtain a confidence limit as the integral of Equation \ref{eqn:pdf} over $\tau>0$ is not finite for $k=1$. This yields a timescale of $3.0^{+12.7}_{-1.9}$~kyr for the class I stage.

%Therefore we can only estimate the outburst rate during the class I phase, using the same arguments from Section \ref{sec:outrate} we determine a timescale of $3.0^{+12.7}_{-1.9}$kyr. Therefore, the lack of outbursts in the class II sample of \citeauthor{2013Scholz} only yields a lower limit of 14 kyr for the outburst rate in class II YSOs. We can also use this to determine the outburst rate during the class I phase, using the same arguments from Section \ref{sec:outrate} we determine a timescale of $3.0^{+12.7}_{-1.9}$kyr. 

\begin{figure}
	\resizebox{\columnwidth}{!}{\includegraphics{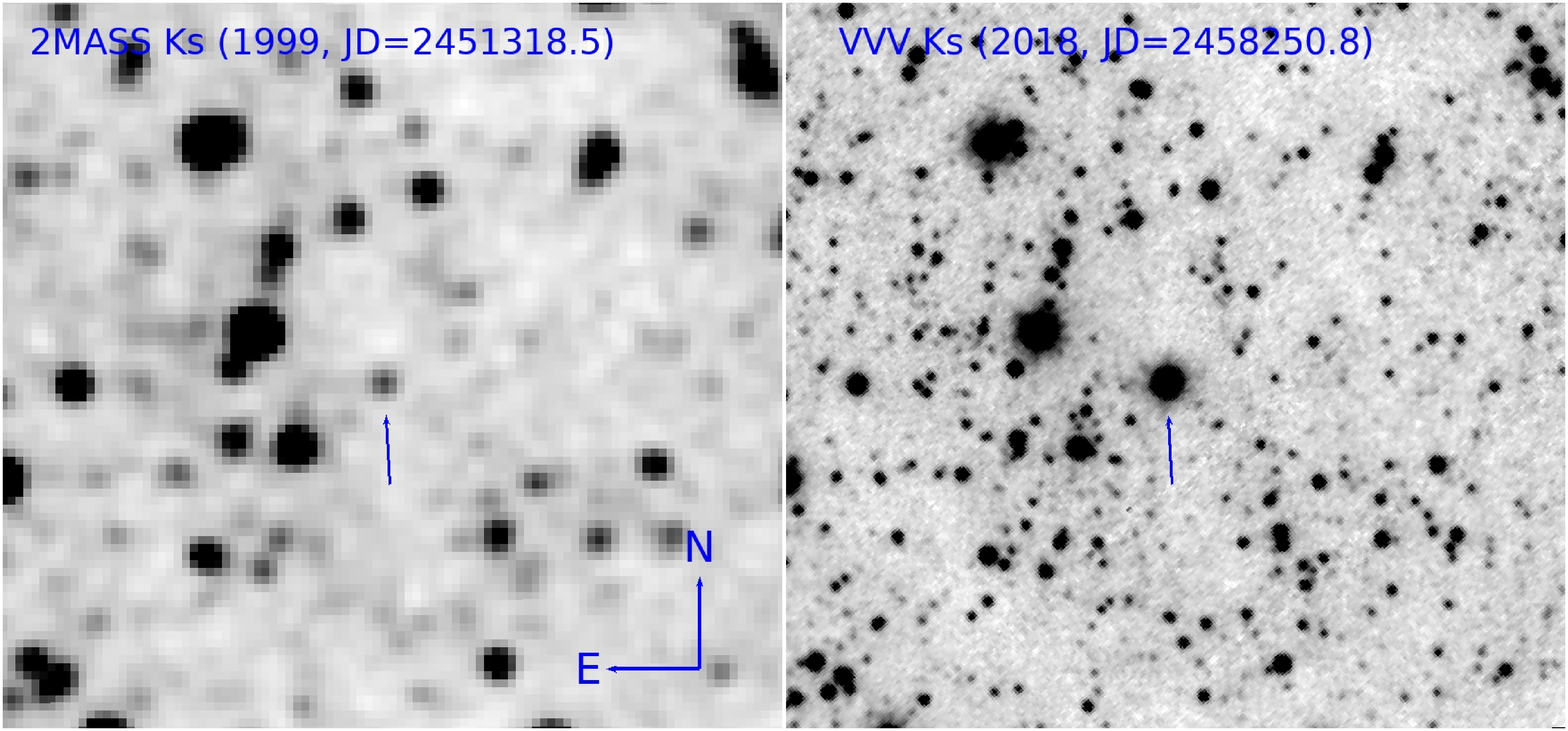}}\\
	\resizebox{\columnwidth}{!}{\includegraphics{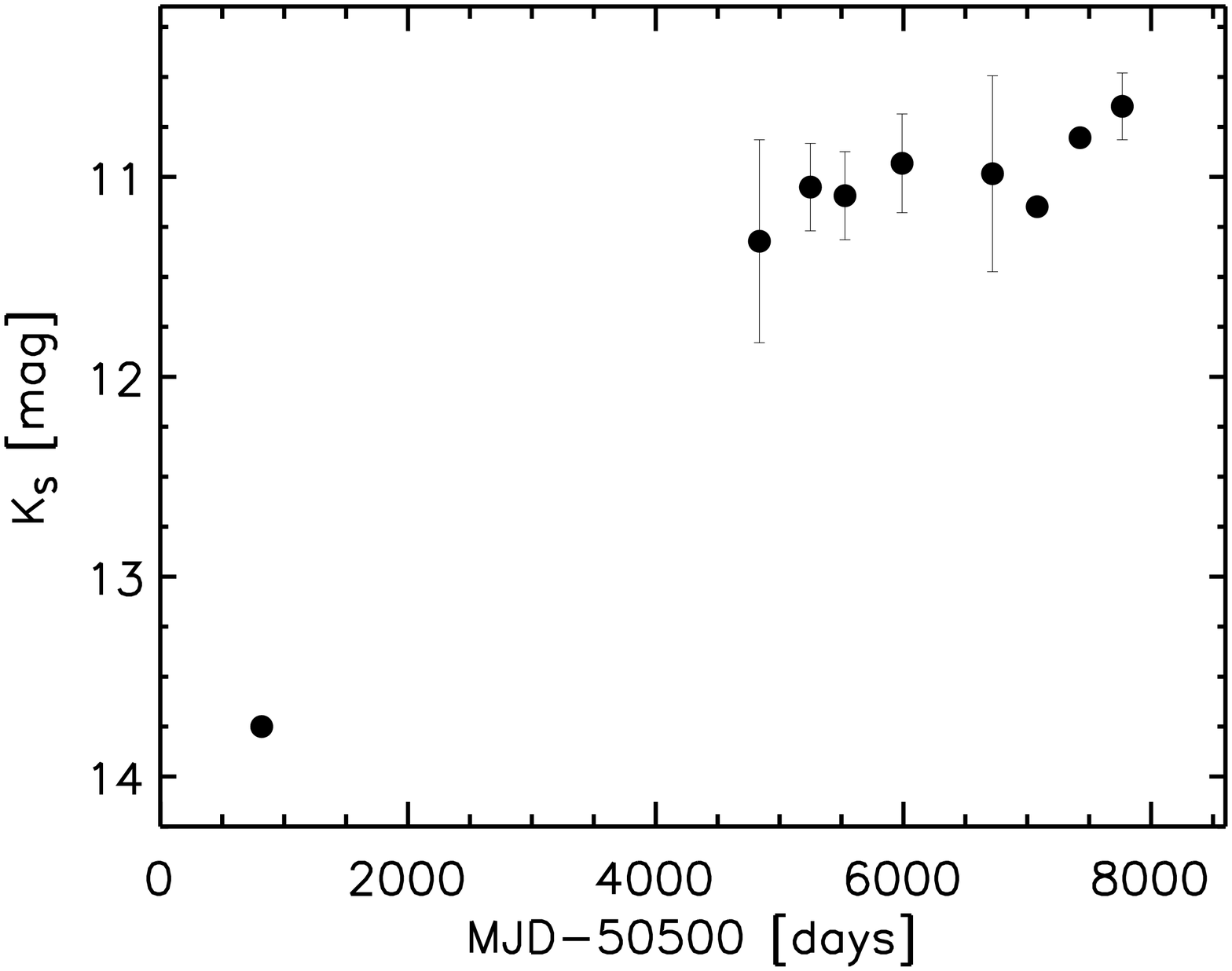}}\\
	\resizebox{\columnwidth}{!}{\includegraphics{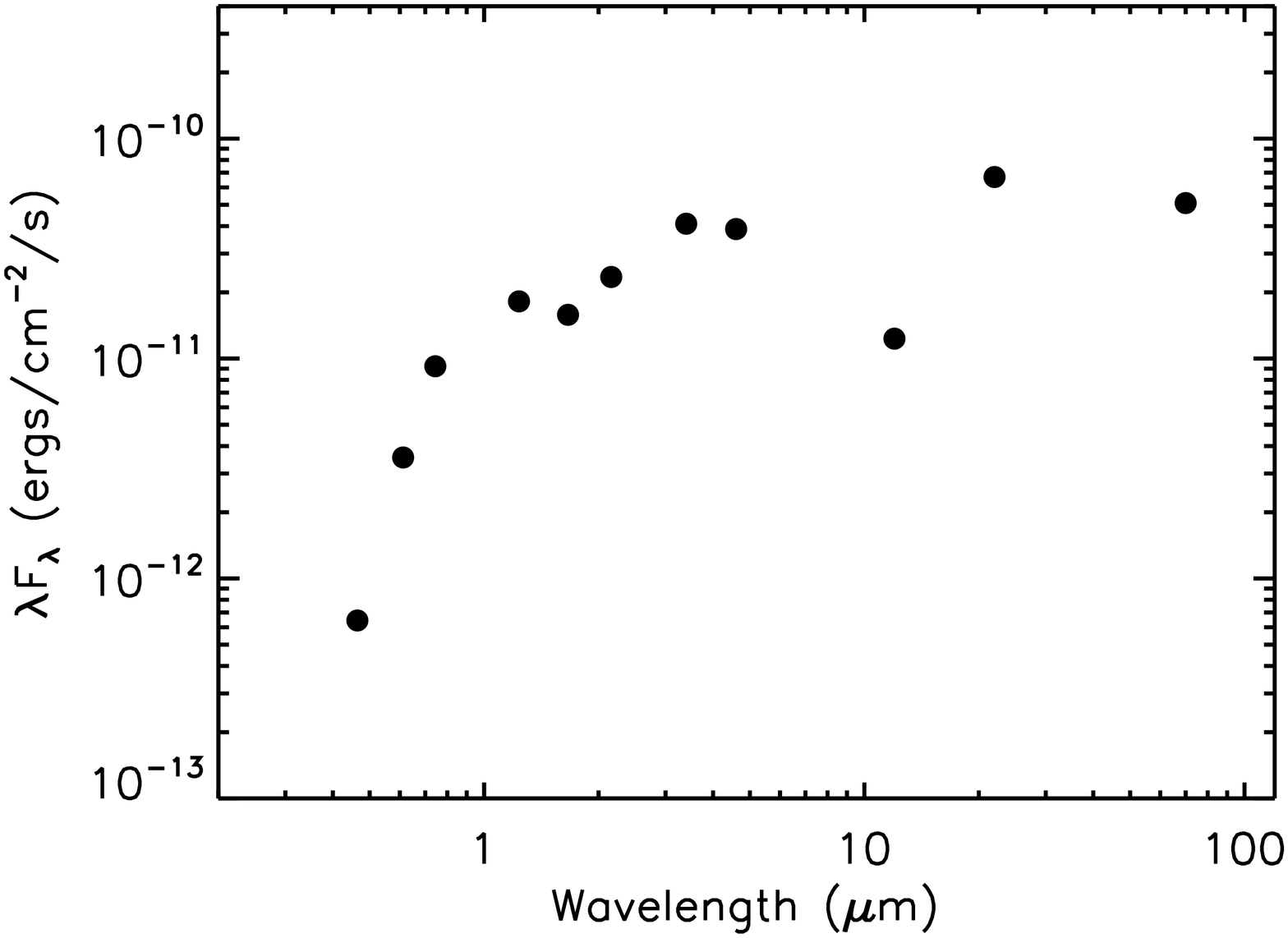}}
    \caption{(top) Comparison of $1.5\arcmin\times1.6\arcmin$ $K_{\rm S}$ images for SFW13 YSO1 from 2MASS (left) and VVVX (right). In both images the YSO is marked by a blue arrow. (middle) $K_{\rm S}$ light curve of SFW13 YSO1 arising from 2MASS, VVV and VVVX. The object is close to the saturation limit in the VVV and VVVX images and thus suffers from non-linearity effects, making the individual data points unreliable. Given this, for VVV and VVVx, we averaged the photometry for each observing campaign. The error bars mark the standard deviation from the average. (bottom) SED of SFW13 YSO1.}
    \label{fig:sfw1}
\end{figure}

In sample B \citet{2013Scholz} find two outburst candidates, 2MASS J16443712-4604017 (hereafter SFW13 YSO1) and 2MASS J15111357-5902366 (hereafter SFW13 YSO2). Using the multi-epoch $K_{\rm s}$ photometry from the VISTA Variables in the Via Lactea survey and its extension \citep[VVV and VVVx respectively, see e.g.][]{2012Saito,2018Bica} we determined that SFW13 YSO2 showed high-amplitude recurrent variability over the 2010-2018 period, whilst SFW13 YSO1 has remained at a bright state since 2010 until at least the latest epoch of VVVX (May 2018, see Figure \ref{fig:sfw1}), making the YSO a true eruptive variable. Using the \citet{2014Koenig} classification criteria, we determined that SFW13 YSO1 is a class II YSO from its {\it WISE} colours. However the object shows a rising SED (Figure \ref{fig:sfw1}) with an spectral index, $\alpha=0.12$ that is more consistent with a class I YSO. Thus, we assume the latter classification for the object. 

An analysis of near- to mid-infrared colours of candidate YSOs from red objects in \citet{2008Robitaille} shows that $\simeq40$\% can be classified as class II YSOs. Assuming the same fraction for sample B in \citet{2013Scholz} implies that around 1600 objects are class II YSOs. Since the only outburst in sample B, SFW13 YSO1, is more likely a class I YSO, we can only estimate that the recurrence timescale in the class I stage, using the $1/\tau$ prior, of $6.0^{+25.3}_{-3.9}$kyr. Similarly to the results of sample A, we can only say that the outburst recurrence timescale during the class II stage is longer than $\simeq$8 kyr.

 %II phase is longer than 8 kyr. At the same time, we determine a timescale of $6.0^{+25.3}_{-3.9}$kyr for the class I phase.

The lack of class II outbursts in the samples of \citet{2013Scholz} agrees with the results from our work, i.e. given the outburst rate determined by us we would not expect to observe an outburst for a sample of a few thousand class II YSOs over a 5 yr baseline.

The analysis of the \citet{2013Scholz} yields another important result, as samples A and B show that during the class I stage, outbursts due to variable accretion appear to be $\approx$10 times more frequent than during the class II phase.

%From the analysis of the \citet{2013Scholz} sample we can only estimate that the outburst rate in class II YSOS is longer than 12-14 kyr. This still agrees with the result from this work, as we do not expect the outburst rate to be shorter than 100 kyr.

%yields lower limits for the outburst rate in class II YSOS, with this rate being longer than 14 kyr. This agrees with our estimate, as we do not expect the outburst rate to be shorter than 55 kyr. Most importantly, samples A and B show that during the class I phase, outbursts due to variable accretion appear to be $\approx$10 times more frequent than during the class II phase.

%Importantly, our study shows that although less frequent than during the embedded phase, outbursts during the class II stage still occur and are can thus have an important effect on planet formation and evolution.

\subsubsection{Theoretical estimate}

\citet{2014Bae} find that gravitational instability(GI)-induced spiral density waves can heat the inner disc and trigger magneto-rotational instabilities (MRI) that will lead to accretion outbursts (or GI+MRI mechanism). They find that this process is more efficient at earlier stages when the envelope dominates the disc, with the number of outbursts decreasing as the star approaches the post-infall phase. Based on figure 3 of \citet{2014Bae}, \citet{2015Hillenbrand} estimate an outburst rate of $8\times10^{-5}$~year$^{-1}$~star$^{-1}$ during the class I stage, and $3\times10^{-6}$~year$^{-1}$~star$^{-1}$ during the class II stage, or a recurrence timescale of 12.5 and 333 kyr respectively. The value for the class I stage is consistent with our class I estimates. The theoretical estimate for the class II stage agrees with our results in the fact that is much longer than that of the class I stage, and that outburst occur over scales longer than 100 kyr. However, it is still longer than the recurrence timescale from our study, and is formally rejected at a 95.3\% confidence level. 

\subsubsection{Other observational evidence}

The recurrence timescale from our study is larger than the time between ejection events of $\simeq1000$~years determined from the observed gaps between H$_{2}$ knots along jets from YSOs \citep{2012Ioannidis, 2016Froebrich, 2018Makin} and the 100 year timescale observed in MNors by \citep{2017Contreras_a}. However this is not surprising as H$_{2}$ jets likely trace the accretion related events during the class I phase \citep{2012Ioannidis} and that the young eruptive variables in \citet{2017Contreras_a} are biased towards class I YSOs. 

\citet{2019Fischer} studied the large amplitude mid-infrared variability for a sample of 319 protostars (YSOs with class designations 0, I or flat) in Orion, by comparing {\it Spitzer} versus {\it WISE} photometry. Based on the recovery of two outbursts over the 6.5 yr baseline \citet{2019Fischer} determine an outburst rate of 1000 yr with a 90\% confidence interval of 690 to 40300 yr. This agrees with our results that suggest that outbursts are more frequent towards YSOs at earlier evolutionary stages the class II stage.

\subsubsection{Gaia17bpi and the rate from the {\it Gaia} alerts}

At the time of writing, \citet{2018Hillenbrand}, based on photometry from {\it Gaia}, reported the discovery of a new YSO outburst that has the characteristics of the most powerful events. The outburst occurred around 2014 and is still ongoing, so it may well be a long-lasting event. The YSO prior to outburst was faint at $r=22$~mag and it is difficult to classify the source from its SED, but it could correspond to a class II YSO \citep[see figure 2 of][]{2018Hillenbrand}.  

The discovery of a potential class II YSO outburst over the 3 year baseline of {\it Gaia} does not contradict our results. Given an outburst rate of 112 kyr, to observe an outburst over a 3 year baseline would imply that we are observing a sample of approximately 40000 YSOs. This is conceivable if we take into account that there is a large population of YSOs that are too faint for current and past surveys. The number of YSOs increases as we go towards fainter magnitudes \cite[see e.g.][]{2017Contreras_a} and Gaia17bpi itself was in fact unknown prior to its outburst. \citet{2018Hillenbrand} followed up the YSO as a {\it Gaia} alert \citep{2013Hodgkin} was triggered on a source that was found in proximity of a star forming region.

The outburst of Gaia17bpi is remarkable as the analysis of \citet{2018Hillenbrand} shows that the outburst started first at mid-infrared wavelengths, with the optical outburst occurring at least a year later. This is consistent with instabilities the lead to an ``outside-in'' type outburst, which is predicted by the outburst mechanisms that could explain outbursts during the class II stage (see Section \ref{sec:out_mec}).

%\citet{2015Hillenbrand} provide arguments to constrain the rate of powerful, long-lasting accretion events. Based on their analysis, and assuming a detection efficiency of $\epsilon=1$, our results imply and outburst rate of  $r=10/(55.6\times15404)$ or $r=1.16\times10^{-5}$~year$^{-1}$~star$^{-1}$, with a 90$\%$ confidence interval of [$7.2\times10^{-6}$,$1.98\times10^{-5}$] outbursts per star per year. 

\section{Caveats}\label{sec:caveats}

%\subsection{caveats}

%The outburst rate obtained in our study is subject to a number of caveats. We have already discussed the issues with the classification of YSOs into different evolutionary classes in Section \ref{ssec:ysoc}. In the following we will discuss additional issues.

%\subsubsection{Outburst mechanism}

%We have mentioned that planet-induced thermal instability and binarity as the likely mechanisms driving large accretion events. However, this has implications for our interpretation of the outburst rate. As noted by \citet{2013Scholz}, the fact that outbursts are triggered by different mechanisms that might not apply to every YSO, would imply that the frequency of bursts is highly variable amongst young stars and it would not be valid to extrapolate from our sample of YSO bursts.

%\begin{description}
\subsection{Contamination by other classes of variable stars}\label{ssec:mimic}

The process of planet formation is more likely to be affected by long-lasting rather than short-term accretion events. In this sense we needed to be careful not to include objects that go through sudden episodes of large changes in the accretion rate, but with a short duration. For example, V8 is the known short-term eruptive YSO ASASSN13-db \citep{2017Sicilia}. When inspecting its long-baseline light curve the object did not resemble a long-lasting outburst, but instead appeared to be irregular, i.e. going from bright to faint states repetitively across time. As a result we are confident that  similar objects that suffer short-term accretion events were not included in our final sample of long-lasting YSO outbursts.

The classification of high-amplitude variable stars as showing long-lasting outbursts, based only on the photometric properties of the sample, also has the risk of selecting objects with light curves that can mimic an accretion-related outburst, especially if the light curve sampling is sparse (as discussed in Section \ref{ssec:longvar}). For example is not hard to imagine that sparse sampling of objects brightening from high-amplitude, long-lasting eclipses by inhomogeneities at long distances in the accretion disc (RW Aur, AA Tau) or a precessing disc in a binary system \citep[e.g. KH 15D,][]{2018Aronow} could resemble a YSO outburst. 
 
 Contemporaneous multi-colour monitoring and/or spectroscopic follow-up are usually used to determine that the dramatic change in the brightness of the system is due to large changes in the accretion rate. The sample of 6 long-lasting YSO outbursts that were used to determine the outburst rate in Section \ref{sec:outrate} contains 3 known YSO outbursts that have been confirmed as such via spectroscopic follow up (V2, V10, V36). Hence in this work we have identified three previously unknown outbursting YSOs that are included in our calculation. Of these, V51 has literature data  confirming that this object is an eruptive YSO (see Appendix \ref{app:av51}). Recent spectroscopic monitoring shows that V4 has the spectroscopic characteristics of objects at the high accretion rates of long-lasting YSO outbursts (Contreras Pe\~{n}a et al. 2019, in prep.). V9 is the only object lacking such an spectroscopic confirmation. 

However, as discussed in Section \ref{ssec:longvar} we tried to avoid including objects where identification as long-lasting YSO outburst was uncertain due to bad sampling and/or lack of additional data from the literature. As a result we are likely excluding objects that could mimic strong YSO outbursts.

\subsection{Non-YSO contamination}\label{ssec:non_yso}

 The sources of contamination in the classification of YSOs arises from red extragalactic sources, such as obscured active galactic nuclei (AGNs), and Galactic sources, mainly reddened giant stars or dust-enshrouded asymptotic giant branch (AGB) stars \citep[see e.g][]{2008Robitaille, 2013Povich}. However, we do not expect the contamination in our sample to be high, as most of our objects arise from samples where contaminants had already been dealt with. For example, \citet{2016Marton} estimate a less than 1$\%$ contamination in their sample of YSOs, whilst this number is not larger than 2$\%$ in the samples of \citet{2012Megeath} and \citet{2009Gutermuth}.  

We could expect a higher percentage of contamination from the sample of objects that are classified using ALLWISE photometry. \citet{2014Koenig} establish that a contamination of $7\%$ is expected in objects classified as class II YSOs from ALLWISE photometry. If we assume that objects classified from ALLWISE or 2MASS photometry alone (6419 objects) suffer this high contamination, and that objects selected from other methods suffer a 1\% contamination, then we expect $\simeq492$ or 3.5\% of objects are contaminants in our sample. 

\subsection{Mis-classification}\label{ssec:ysomisc} The likely evolutionary stage for the majority of our sample was determined using the \citet{2014Koenig} criteria. Inspection of their figure 5 shows that the area in the colour-colour plane that defines the class II YSOs could be contaminated by objects at earlier evolutionary stages, i.e. flat-spectrum and class I sources. Given the much higher outburst rate of lass I sources compared with class II (Section \ref{sssec:scholz13}), even a small percentage of misclassification would lead to the detection of class I YSOs in our outburst sample, an effect that we observed in our results (see Section \ref{ssec:ysoc}).

To estimate this contamination, we obtained ALLWISE photometry for a sample of YSOs from \citet{2012Megeath} that also have measured values of the spectral index, $\alpha$. Then we applied the \citet{2014Koenig} criteria to determine the likely evolutionary stage of these YSOs. Finally, we compared the latter with the stage determined from $\alpha$. We found that 7\% of objects classified as class II YSOs from the \citet{2014Koenig} criteria are classified as being at earlier evolutionary stages from the value of the spectral index. As we will show in the following, this estimate is consistent with the number of class I YSO outburst that contaminate our sample in Section \ref{ssec:ysoc}

\subsection{Effects of contamination and mis-classification on the outburst rate}\label{ssec:effects}

We can now examine the effect of all of the uncertainties discussed in Sections \ref{ssec:mimic} to \ref{ssec:ysomisc} on our outburst rate. If we first only consider the effect of the lack of spectroscopic confirmation as an accretion outburst for V9 (Section \ref{ssec:mimic}), then our outburst sample reduces from 6 to 5 objects. This implies an increase of the outburst rate, from 112 to 130 kyr.

Considering possible contamination of non-YSOs, our sample reduces to 13589 objects (see Section \ref{ssec:non_yso}). If we further consider a 7\% contamination from class I YSOs (951 objects, see Section \ref{ssec:ysomisc}) then we are left with 12638 class II YSOs. If we use a sample of 6 YSO outbursts, the outburst rate in class II YSOs reduces from 112 to 100 kyr. If we omittV9, the outburst rate increases again to 117 kyr. 

Thus considering all of the possible contamination and exclusion effects, the outburst rate changes to between 100 and 130 kyr. This is a change that is within the 68\% confidence intervals we estimated in Section \ref{sec:outrate}.

We can also use the contamination of class I YSOs to determine another approximate estimate of the outburst rate in the class I stage. In Section \ref{ssec:ysoc} we found that 3 YSO outbursts are more likely to be class I YSOs, if we consider a total sample of 1043 class I YSOs implies an outburst rate of $13^{+15}_{-6}$\,kyr (68 percent confidence, assuming the $1/\tau$ prior). This number agrees remarkably well with that estimated from the \citet{2013Scholz} YSO sample.

\subsection{Selection criteria}

The value for the recurrence timescale derived from our sample of  class II YSO outbursts is subject to some caveats from the initial selection. As we are interested in a statistical characterisation of variability, we should not use variability as a selection criterion. However, is clear that some of the objects in our initial catalogue were initially identified because of their variability, primarily the  FU* and or* classes. However given the need of IR data to identify class II objects, we believe that they would have also have been identified as YSOs in these surveys, regardless of their variability. Hence we left them in the sample.

\section{Outburst mechanism}\label{sec:out_mec}

The results from our analysis show for the first time that the suggestion that the frequency of the outbursts is lower in the class II stage than during the class I stage is correct. This has implications for the physical mechanism likely to be driving the outbursts during class II phase. 

As noted above GI+MRI instabilities become less efficient in disc dominated systems \citep{2014Bae}. Due to GI, and provided that the cooling time is shorter than the dynamical timescale, disc pressure is not able to support the collapsing gas against its own gravity, leading to disc fragmentation \citep[see e.g., ][]{2005Vorobyov, 2011Machida}. The fragments are later transported to the inner disc via gravitational interactions with the spiral arms \citep{2011Machida}, and are likely to trigger mass accretion bursts \citep{2010Vorobyov}. This process is more efficient during the embedded phase. Mass accretion outbursts could occur during the class II phase, but it is not clear whether these fragments survive beyond the embedded stage \citep[see ][]{2014Audard}. \citet{2010Zhu} find that for the lower infall rates of class II objects, layered MRI turbulence might be able to accumulate mass and trigger the MRI outbursts. However, these bursts are likely of lower amplitude and of shorter duration \citep{2014Audard}. 

\citet{2004Lodato} propose that the interaction of a massive planet with the disc can trigger thermal instabilities in the outer disc. In this scenario, the migration of the planet opens up a gap in the disc, and the inner disc is emptied out. Due to the tidal effect induced by the planet, material will pile up at larger radii. When the density reaches a critical value, the thermal instability is triggered. A final possibility is that the gravitational force of a companion star may perturb the disc and thus enhance accretion \citep{1992Bonnell}. This idea is supported by the fact that a number of known eruptive variables are members of binary systems \citep[see e.g., ][]{2007Fedele,2010Reipurth, 2015Caratti}. In some of these, both stars in the system are found to have characteristics of strong accretors \citep[e.g. systems AR6A+B, RNO1B/C, ][]{2003Aspin}.

Therefore, the planet-induced thermal instability and binary models are most likely to explain the existence of large accretion events during the class II stage. However, we note that the fact that more than one mechanism can explain the outbursts has implications for our interpretation of the outburst rate. As noted by \citet{2013Scholz}, if outbursts are triggered by two or more different mechanisms not all of which might apply to every YSO (such as outbursts due to encounters in multiple systems) would imply that the frequency of outbursts is highly variable amongst young stars, and it would not be valid to extrapolate the outburst rate for a single YSO from our estimated class II recurrence timescale.

%\end{description}

%It is hard to quantify the true effects of these caveats on our estimated outburst rate. The effects of contamination are likely to increase the value of the outburst rate, but the effects of misclassification would reduce this value. However these values are still probably contained within the 68 percent confidence interval estimated above. Therefore these caveats do not have a major influence on the conclusions of our study.

\section{Summary}\label{sec:sum}

In this work we compared the all-sky digitised photographic plate surveys provided by SuperCOSMOS with the latest data release from {\it Gaia} (DR2) for a large sample of YSOs. The mean baseline of 55 years implies that we monitored $\simeq800,000$ YSO years.

We retrieved 139 high-amplitude variables with $\Delta R\geq1$~mag. We believe we are complete for amplitudes larger than 2 magnitudes. In Section \ref{sec:sel_hamp} we determined that most of the variable stars are found with amplitudes between 1 and 3 mag, where variable extinction or extreme changes due to hot-spot variability are the likely mechanisms driving the variability in these YSOs. For amplitudes larger than  3 mag we also find objects with apparent long-lasting fading events, likely related to extinction, and objects with irregular variability.  At these larger amplitudes, the latter are more likely explained by short-term accretion events that appear as irregular given the long baseline of the light curves.

YSOs with this extreme variability, but not related to variable accretion, are interesting in their own right. For example, long lasting fading events could be explained by extinction due to inhomogeneities at large distances in the disc, which can relate to planet-formation processes.

Objects where variability could be due to changes in the accretion rate, but with outburst duration that is less than 10 years, were not included in the calculation of the recurrence timescale of outbursts in the class II stage. This is because short duration outbursts are less likely to have an impact on the formation and evolution of protoplanets \citet[see e.g.][]{2017Hubbardb}.

We classified 9 YSOs as showing long-term accretion driven outbursts, with 3 of them representing new discoveries (Section \ref{ssec:outb}). We discussed the possible contamination by class I YSOs and we determine that 3 objects in this group are more likely at a younger evolutionary stage. Therefore, 6 YSOs are left in our final sample of class II YSO outbursts (Section \ref{ssec:ysoc}).

In Section \ref{sec:outrate} we determined, for the first time, that long-term accretion driven outbursts occur during the class II stage of young stellar evolution. The frequency of these events was found at $\tau$=$112^{+68}_{-38}$\,kyr (68 percent confidence). Our results imply that planet formation models must take into account the likely effects that these episodic accretion events can have on the formation and evolution of planetary systems.

Through the analysis of the YSO sample of \citet{2013Scholz} (see Section \ref{sssec:scholz13}) as well as possible contamination of class I YSOs in our sample (Section \ref{ssec:effects}), we have estimated the recurrence timescale of accretion driven outbursts at this earlier evolutionary stage. We found that episodic accretion events are $\simeq$10 times more frequent than during the class II stage, in agreement with theoretical expectations.

\section*{Acknowledgements}

This work has made use of data from the European Space Agency (ESA) mission {\it Gaia} (\url{https://www.cosmos.esa.int/gaia}), processed by the {\it Gaia} Data Processing and Analysis Consortium (DPAC, \url{https://www.cosmos.esa.int/web/gaia/dpac/consortium}). This research has made use of the NASA/ IPAC Infrared Science Archive, which is operated by the Jet Propulsion Laboratory, California Institute of Technology, under contract with the National Aeronautics and Space Administration. Funding for the DPAC has been provided by national institutions, in particular the institutions participating in the {\it Gaia} Multilateral Agreement. The contributions of C.C.P. and T.N. were funded by a Leverhulme Trust Research Project Grant and of S.M. through a Science and Technology Facilities Council (STFC) studentship. We are grateful to N. Hambly for his help in understanding the SSS database.

%The Acknowledgements section is not numbered. Here you can thank helpful colleagues, acknowledge funding agencies, telescopes and facilities used etc.Try to keep it short.

%%%%%%%%%%%%%%%%%%%%%%%%%%%%%%%%%%%%%%%%%%%%%%%%%%



\bibliographystyle{mnras}
%\bibliography{ref_sc.bib} % if your bibtex file is called example.bib


%%%%%%%%%%%%%%%%%%%%%%%%%%%%%%%%%%%%%%%%%%%%%%%%%%

%%%%%%%%%%%%%%%%% APPENDICES %%%%%%%%%%%%%%%%%%%%%

\appendix

\section{Individual notes}\label{app:a}

In the following section we will describe the three YSOs that are previously unknown as eruptive variables. The individual subsections include a description of the photometry and any information that we were able to gather from the literature and that allowed us to include these objects as YSOs displaying long-term outbursts.

\subsection{V4}

V4 (2MASS J02335340$+$6156501) was originally classified as a YSO by \citet{2014Panwar}  based on its near-infrared excess from 2MASS data. There is no additional literature information for this source. The near-infrared $JHK_{\rm s}$ colours from 2MASS indicate that this is likely a class II YSO, hence its inclusion in our sample.

\begin{figure*}
\resizebox{0.8\textwidth}{!}{\includegraphics{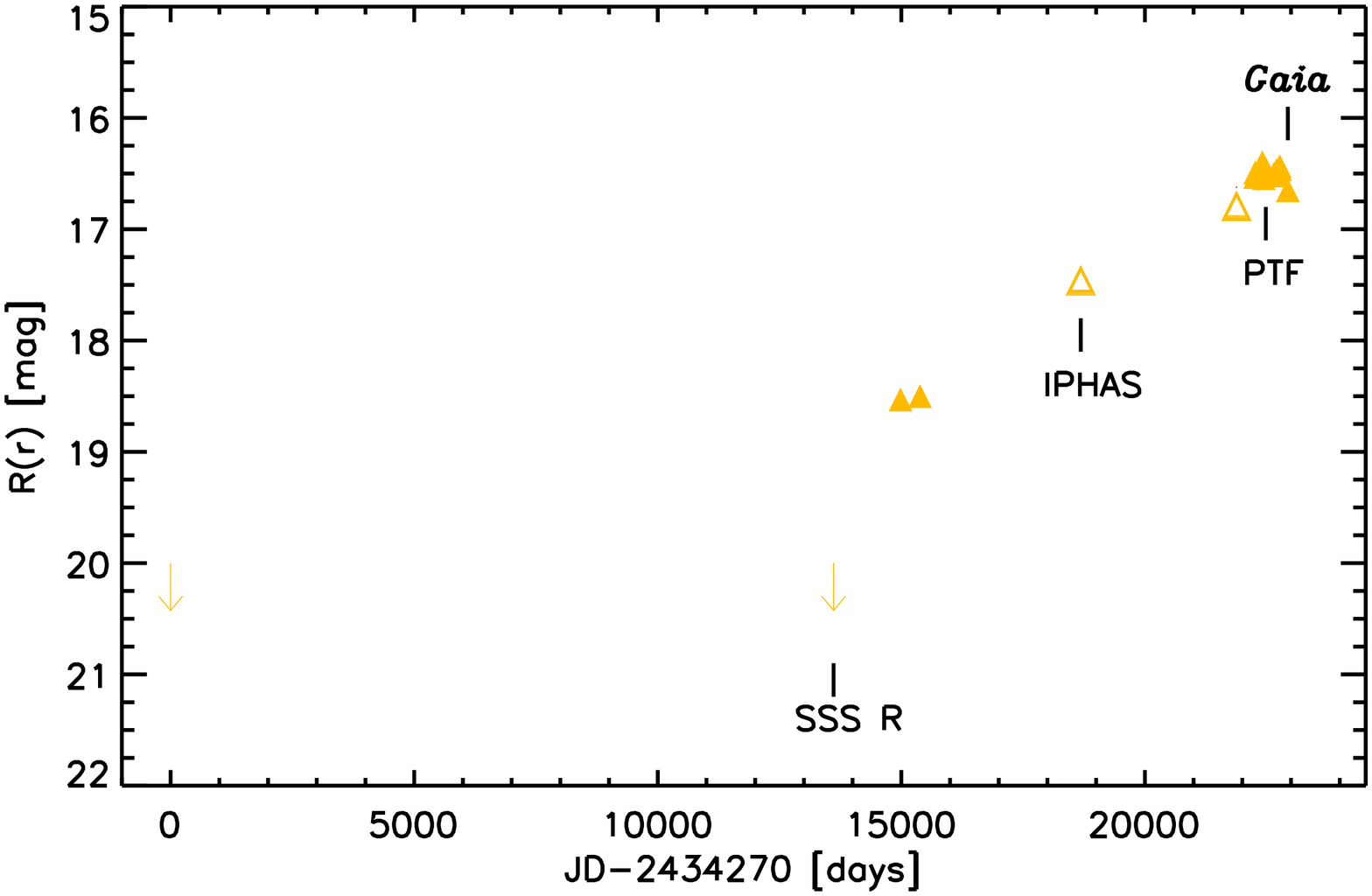}}\\
\resizebox{0.8\textwidth}{!}{\includegraphics{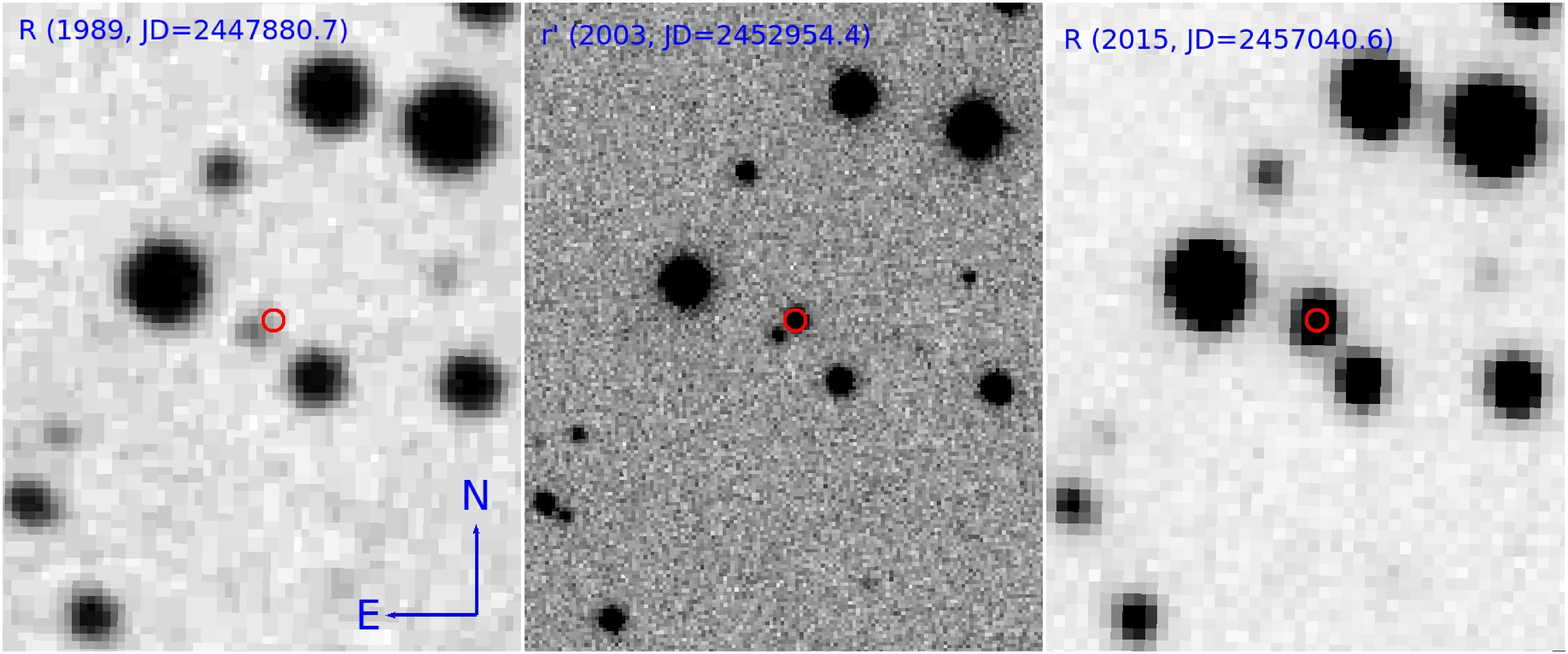}}
\caption{ (top) Light curve incorporating all available data found in the literature for V4. The photometry from different filters has been converted to approximate R-band magnitudes (shown as solid triangles) to help the clarity of the light curve. Sloan r filter magnitudes (open triangles) were not converted. In the plot we mark the approximate epoch of each survey for which we present images in the bottom figure. (bottom) Comparison between the $R$-band SSS image obtained in 1989 (left), sloan $r$ IPHAS image from 2003 (middle) and the latest PTF $R$ image from 2015 (right). In the figure all images have a size of $1\arcmin\times0.8\arcmin$.}
  \label{fig:hist_v4}
\end{figure*}

The classification as a high-amplitude variable for this source results from its {\it Gaia} DR2 magnitude being brighter than all of the SSS bands. The object is not visible in the images of  the 2 epochs of $R$-band from SSS. Although the SSS catalogues yield a detection with $R=19.3$ this in fact corresponds to a star located 2\arcsec south west from V4. The variable star becomes visible in the $B$ and $I$ SSS observations and continues to get brighter as confirmed from IPHAS, Pan-STARRS and PTF observations (see Figure \ref{fig:hist_v4}).

\subsection{V9}

\begin{figure*}
\resizebox{0.8\textwidth}{!}{\includegraphics{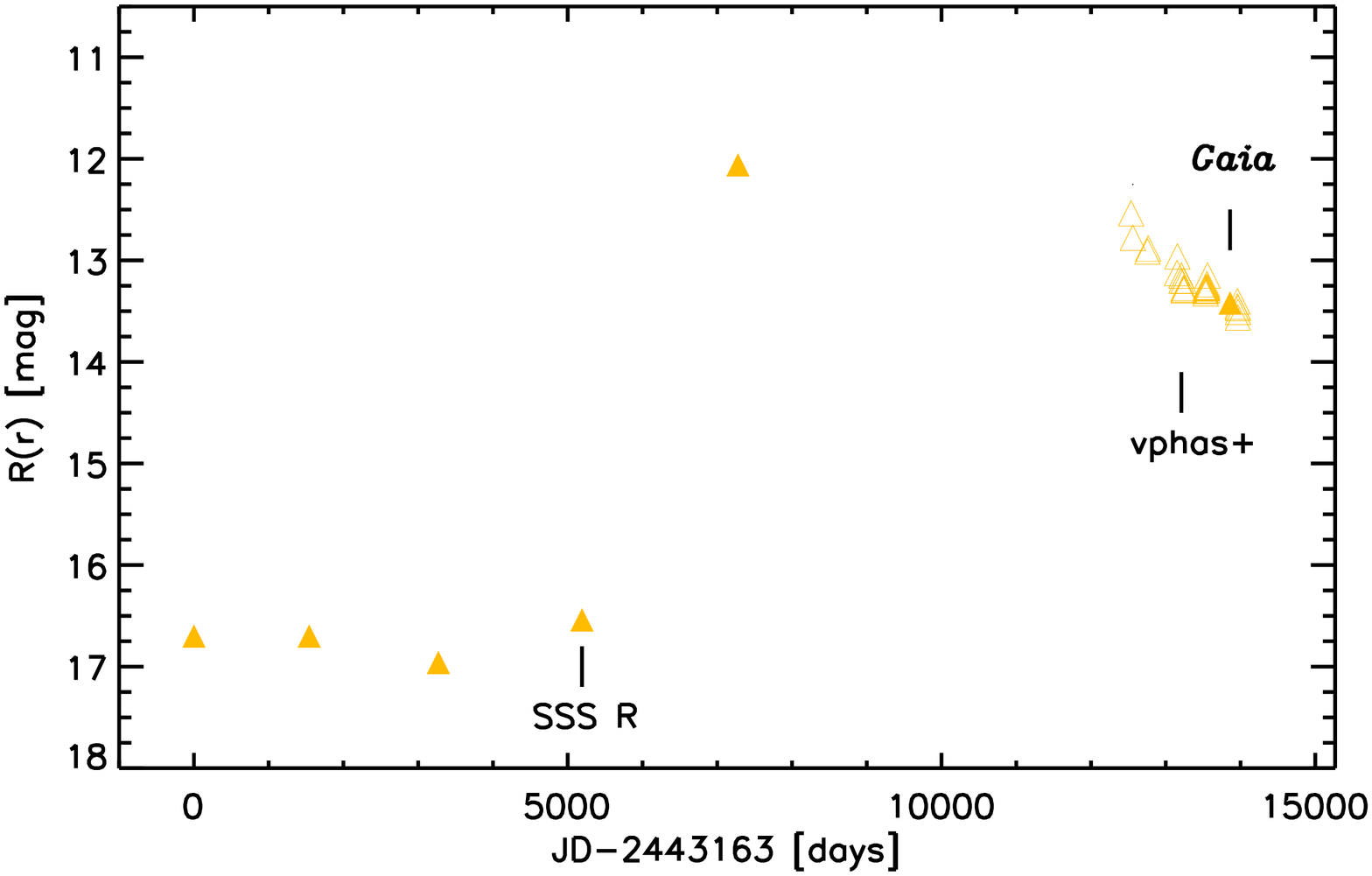}}\\
\resizebox{0.8\textwidth}{!}{\includegraphics{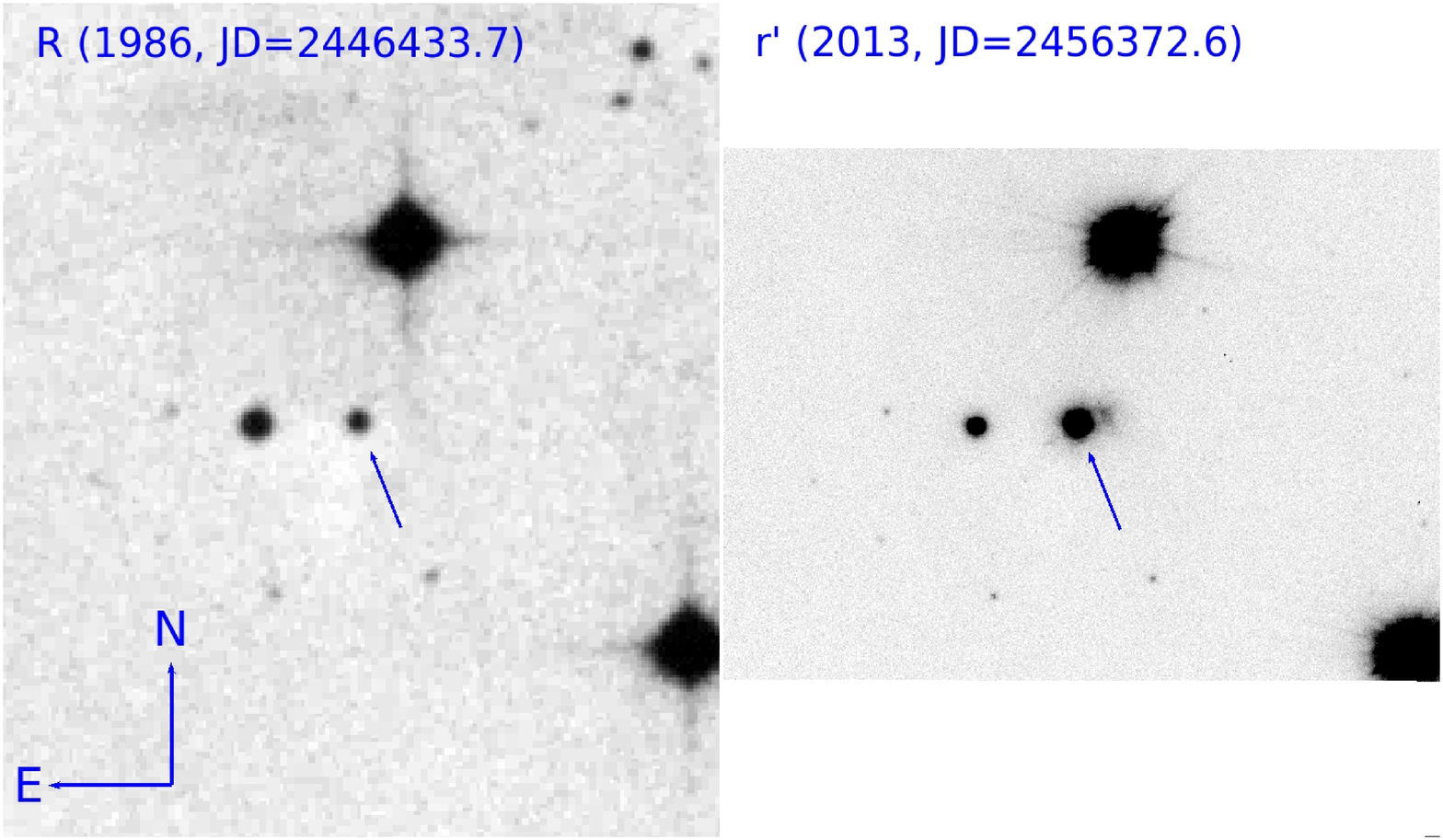}}
\caption{ (top) Light curve incorporating all available data found in the literature for V9. The photometry from different filters has been converted to approximate R-band magnitudes (shown as solid triangles) to help the clarity of the light curve. Sloan r filter magnitudes (open triangles) were not converted. In the plot we mark the approximate epoch of each survey for which we present images in the bottom figure. (bottom) Comparison between the second epoch of $R$-band SSS $2.5\arcmin\times2.3\arcmin$ image obtained in 1986 (left), and the VPHAS$+$ sloan $r$ $1.7\arcmin\times2.3\arcmin$ image from 2013 (right).}
  \label{fig:hist_scv59}
\end{figure*}

V9 (2MASS J08410676-4052174)  is an H$\alpha$ emission line star \citep{1994Pettersson}, selected from the \citet{2016Marton} catalogue and classified based on its WISE colours. This object was selected beacuse its {\it Gaia} magnitude appeared brighter than expected compared with all of the 4 SSS bands. Photometry from Vizier confirms this is as a long-term outburst as the object appears to be faint in all SSS epochs with an eruption likely occurring between the second R epoch from SSS and DENIS observations. The object now appears to be slowly declining, as shown by APASS, VPHAS+ observations and the light curve from the Bochum Galactic disc survey \citep{2015Hackstein}. Figure \ref{fig:hist_scv59} shows the light curve of V9 as well as the comparison between SSS and VPHAS$+$ images of the source.

\subsection{V51}\label{app:av51}
\begin{figure*}
\resizebox{0.8\textwidth}{!}{\includegraphics{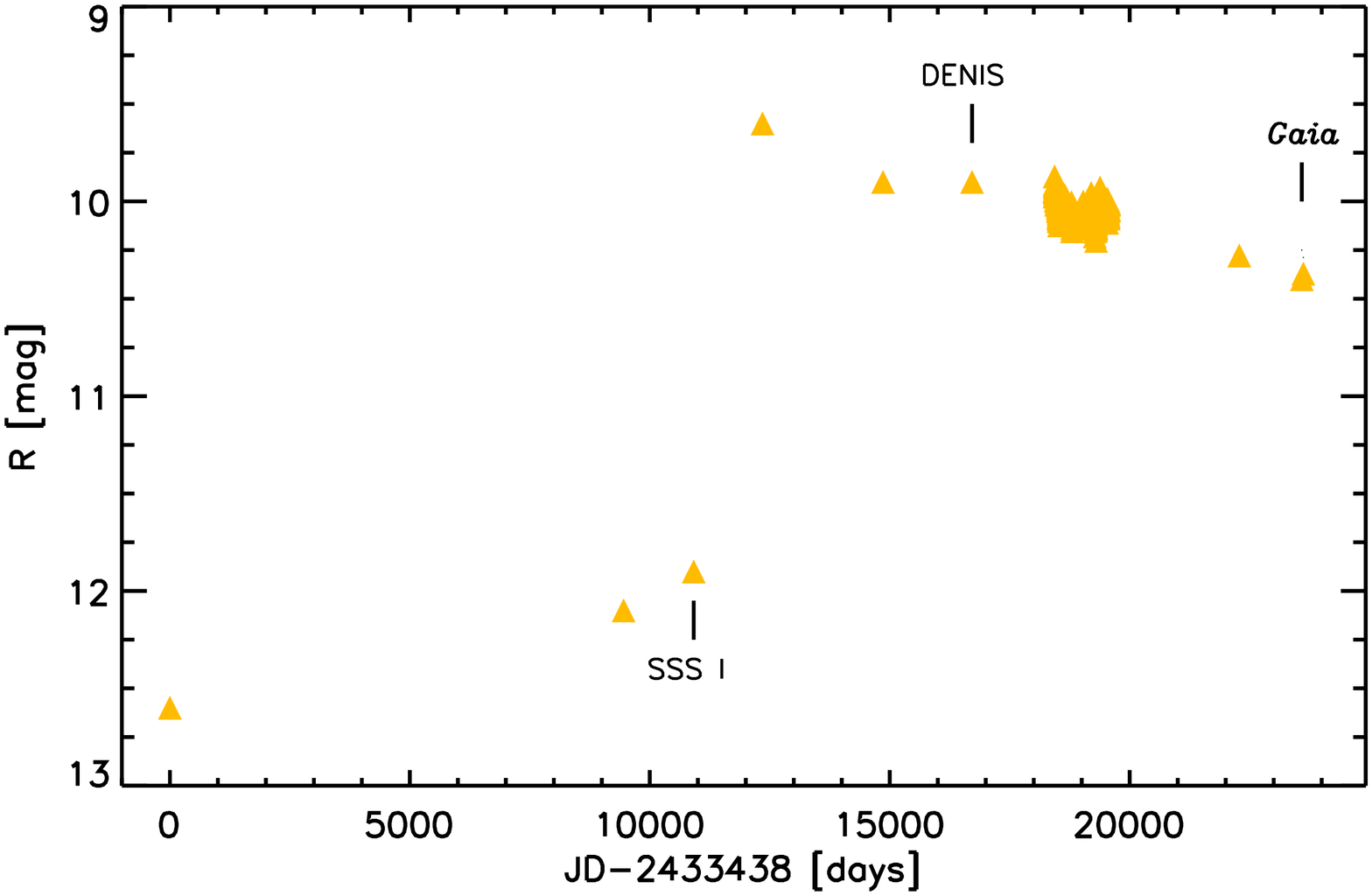}}\\
\resizebox{0.8\textwidth}{!}{\includegraphics{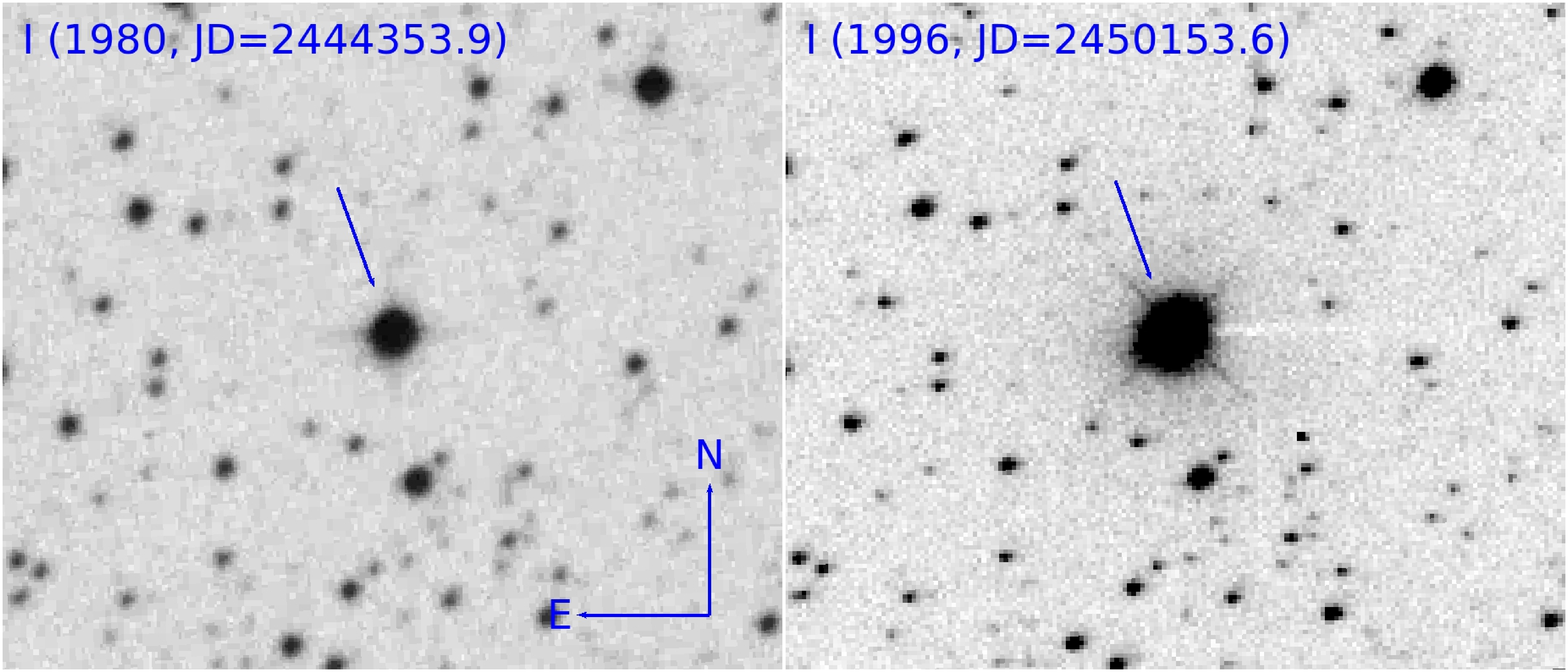}}
\caption{ (top) Light curve incorporating all available data found in the literature for V51. The photometry from different filters has been converted to approximate R-band magnitudes (shown as solid triangles) to help the clarity of the light curve. Sloan r filter magnitudes (open triangles) were not converted. In the plot we mark the approximate epoch of each survey for which we present images in the bottom figure. (bottom) Comparison between the $I$-band SSS image obtained in 1980 (left), and the DENIS $I$ image from 1996 (right). The images in the figure have a size of $2.5\arcmin\times2.9\arcmin$.}
  \label{fig:hist_v51}
\end{figure*}

V51 (WRAY 15$-$488) is an H$\alpha$ emission line star \citep{1966Wray}, which we selected from the \citet{2016Marton} catalogue and classified based on its WISE colours. The object is part of the 139 high-amplitude variable star catalogue as the $R$-band epochs of SSS and {\it Gaia} DR2 are brighter than the earlier B and I SSS observations. Photometry collected from Vizier confirms its nature as a long-term outburst as the object becomes $\simeq$2.5 magnitudes brighter between 1980 (I-band SSS) and 1984 (first epoch in $R$ from SSS) and has remained at approximately the same level ever since, as confirmed by data from the All Sky Automated Survey \citep[ASAS,][]{2002Pojmanski} and SkyMapper (see Figure \ref{fig:hist_v51}).  Evidence that the object was at a quiescent state when it was classified as an H$\alpha$ emission line star by \citet{1966Wray} arises from the catalogue of \citet{1970Wackerling} which contains the data from \citet{1966Wray} and where V51 is listed with V$=13.1$, approximately 3 magnitudes fainter than the current brightness of the object. 

This object has spectroscopic observations that confirm its nature as an eruptive YSO. \citet{2010Carmona}, based on high resolution FEROS optical spectra observed in 2004, find that the object shows an H$\alpha$ P Cygni profile, with an spectrum consistent with that of an F-type giant star. The latter classification is characteristic of eruptive YSOs during outburst \citep{2018Connelley}. Previous spectra, also observed during outburst, from \citet{2006Suarez} and \citet{1992Gregorio} are also consistent with an F-type spectral type. V51 also shows broad Na I D $5890$\AA~absorption \citep{2006Suarez}, another characteristic of YSO outbursts, as well as Li I absorption \citep{1992Gregorio}, evidence of the youth of the system. Remarkably during these spectra \citep[which are observed at earlier epochs of the outburst than that of ][]{2010Carmona} the H$\alpha$ line is seen purely in emission.

\section{Light curves}

\section{Vizier Tables}\label{sec:vizier}

This appendix contains the list of surveys used to add photometry to the light curves of our high-amplitude variable YSOs.

\begin{table*}
\centering
\caption{List of surveys that are part of the analysis of high-amplitude variables found in our study.}
\label{tab:vizier}
\begin{tabular}{lll} % four columns, alignment for each
\hline
Vizier ID  & Title & Reference \\
\hline
V/73A/catalog & Third Catalog of Emission-Line Stars of the Orion Population & \citet{c1988Herbig}\\
III/177/catalog & Observations of Emission-line Stars in the Orion Region I. The Kiso Area A-0904 & \citet{c1989Wiramihardja}\\
B/denis/denis & DENIS: A Deep Near-Infrared Survey of the southern sky &\citet{1994Epchtein}\\
J/AJ/118/1043/tables & A photometric catalog of Herbig Ae/Be stars...UX Orionis stars. & \citet{c1999Herbst}\\
J/AJ/119/3026/table1 & Circumstellar disc candidates identified in the Orion nebula cluster flanking fields. & \citet{c2000Rebull}\\
J/AJ/121/3160/table4 & Near-infrared photometric variability of stars toward the Orion A molecular cloud. & \citet{2001Carpenter}\\
J/AJ/124/1001/table5 & Near-infrared photometric variability of stars toward the Chamaeleon I molecular cloud. & \citet{c2002Carpenter}\\
J/AJ/123/387/phot & A photometric study... $\lambda$ Orionis star-forming region. &\citet{c2002Dolan} \\
J/AJ/123/304/table1 & The young cluster IC 5146 & \citet{c2002Herbig}\\
II/264/var & The All Sky Automated Survey. Catalog of Variable Stars & \citet{2002Pojmanski}\\
2mass-psc  & 2MASS All-Sky Catalog of Point Sources & \citet{c2003Cutri}\\
J/AJ/125/1537/table2 & Deep near-infrared observations... in Orion Molecular Clouds 2 and 3 & \citet{c2003Tsujimoto}\\
J/AN/325/705/table2 & Optical and infrared photometry... $\sigma$ Ori cluster & \citet{c2004Beijar}\\
J/A+A/417/557/table4 & A Rotational and Variability Study of a Large Sample of PMS Stars in NGC 2264 & \citet{c2004Lahm}\\
J/AJ/128/2316/table4 & The low-mass population of Orion OB1b. I. The $\sigma$ Ori cluster & \citet{c2004Sherry}\\
J/AJ/128/1684/table1 & The initial mass function and young brown dwarf candidates in NGC 2264. I. & \citet{c2004Sung} \\
J/AJ/129/829/table1 & The T Tauri star population of the young cluster NGC 2264. & \citet{c2005Dahm}\\
J/MNRAS/356/89/catalog & Membership, binarity and accretion... $\sigma$ Ori cluster & \citet{c2005Kenyon}\\
II/271A/patch2 & TASS Mark IV Photometric Survey of the Northern Sky & \citet{c2006Droege}\\
J/ApJ/662/1067/sigori & A Spitzer Space Telescope study of disks in the young $\sigma$ Ori cluster & \citet{c2007Hernandez} \\
J/MNRAS/375/1220/ngc2264 & Empirical isochrones and relative ages for young stars... NGC2264 & \citet{c2007Mayne} \\ 
J/MNRAS/375/1220/sigori & Empirical isochrones and relative ages for young stars... $\sigma$ Ori & \citet{c2007Mayne} \\ 
II/319/gcs9 & UKIRT Infrared Deep Sky Survey (UKIDSS) Release 9 & \citet{2007Lawrence}\\
II/316/gps6 & UKIRT Infrared Deep Sky Survey (UKIDSS) Galactic Plane Survey (GPS) Release 6 & \citet{c2008Lucas}\\
J/ApJS/183/261/table2 & A multi-color optical survey of the Orion nebula cluster. I. The catalog. & \citet{c2009DaRio}\\
J/ApJS/191/389/table2 & Precision photometric monitoring of very low mass $\sigma$ Orionis cluster members & \citet{c2010Cody}\\
J/ApJ/726/18/stars & Disk evolution in W5: intermediate-mass stars at 2-5 Myr & \citet{c2011Koenig}\\
J/ApJ/733/50/table2 & YSOVAR: ...mid-infrared photometric monitoring of the ONC & \citet{c2011Morales}\\
J/A+A/525/A47/tablec2 & U-band study of the accretion properties in the $\sigma$ Ori star-forming region & \citet{c2011Rigliaco}\\
J/AZh/88/143/oricat & Proper motions of stars in the region of the Great Nebula in Orion & \citet{c2011Vereshchagin}\\
J/A+A/548/A79/omc-var & INTEGRAL-OMC optically variable sources & \citet{c2012Alfonso}\\
J/ApJ/751/22/table3 & A significant population of candidate new members of the $\rho$ Ophiuchi cluster & \citet{c2012Barsony}\\
J/ApJ/752/59/members & The low-mass stellar population in L1641...stellar initial mass function & \citet{c2012Hsu} \\
II/313/table3 & The Palomar Transient Factory photometric catalog 1.0 & \citet{c2012Ofek}\\
J/ApJ/754/30/youth & New isolated planetary-mass objects...$\sigma$ Orionis cluster. & \citet{c2012Pena} \\
J/MNRAS/434/806/catalog & Pre-main-sequence isochrones. II. Revising star and planet formation time-scales. & \citet{2013Bell}\\
J/ApJS/209/28/table2 & The MYStIX wide-field near-infrared data: optimal photometry in crowded fields & \citet{c2013King}\\
II/321/iphas2 & The Second Data Release of the INT Photometric H-Alpha Survey...  & \citet{c2014Barentsen}\\
J/A+A/564/A29/catalog & Orion Revisited: II. The foreground population to Orion A & \citet{c2014Bouy}\\
J/ApJ/794/36/table2 & A spectroscopic census in young stellar regions: the $\sigma$ Orionis cluster. & \citet{c2014Hernandez}\\
J/other/RAA/14.1264/table1 & V350 Cep UBVRI long-term photometry & \citet{c2014Ibryamov}\\
I/327/cmc15 & The Carlsberg Meridian Catalogue, final data release (MC15) & \citet{2014Muinos}\\
V/147/sdss12 & The SDSS Photometric Catalogue, Data Release 12 (DR12) & \citet{c2015Alam}\\
J/AJ/149/200/table4 & An optical survey of the partially embedded young cluster in NGC 7129 & \citet{c2015Dahm}\\
II/336/apass9 & AAVSO Photometric All Sky Survey (APASS) DR9 & \citet{c2015Henden}\\
J/ApJS/220/17/table2 & Wide-field infrared polarimetry of the $\rho$ Ophiuchi cloud core & \citet{c2015Kwon}\\
J/AJ/150/132/variables & Near-infrared variability in the Orion Nebula Cluster & \citet{2015Rice}\\
J/A+A/581/A140/tablea2 & The VISTA Orion mini-survey: star formation in the Lynds 1630 North cloud. & \citet{c2015Spezzi}\\
II/349/ps1 & Panoramic Survey Telescope and Rapid Response System (Pan-STARRS) DR1 & \citet{2016Chambers}\\
II/341/vphasp & The VST Photometric Halpha Survey... (VPHAS+) DR2. & \citet{c2016Drew}\\
J/A+A/587/A153/science & VISION - Vienna survey in Orion. I. VISTA Orion A Survey. & \citet{c2016Meingast} \\
II/348/vvv2 & VISTA Variable in the Via Lactea Survey DR2 & \citet{c2017Minniti}\\
\hline
not in Vizier & Northen Sky Variability Survey (NSVS) & \citet{c2004Wozniak}\\
not in Vizier & The Catalina Real-Time Transient Survey (CRTS) & \citet{c2011Djorgovski}\\
not in Vizier & Panoramic Survey Telescope and Rapid Response System (Pan-STARRS) DR2 & \citet{2016Chambers}\\
not in Vizier & The HOYS-CAPS Citizen Science Project & \citet{c2018Froebrich}\\
not in Vizier & The American Association of Variable Stars Observers (AAVSO) & \citet{c2018Kafka}\\
not in Vizier & SkyMapper Southern Sky Survey & \citet{2018Wolf}\\
\hline
\end{tabular}
\end{table*}

%%%%%%%%%%%%%%%%%%%%%%%%%%%%%%%%%%%%%%%%%%%%%%%%%%

% Don't change these lines
\bsp	% typesetting comment
\label{lastpage}
\end{document}